\documentclass[twocolumn,groupedaddress,assymb,amsmath]{revtex4}
\usepackage{}
\usepackage{graphicx}
\usepackage{amssymb}
\usepackage{amsmath}

\usepackage{extarrows} 
\usepackage[bookmarks=false]{hyperref}
\hypersetup{colorlinks=true,citecolor=blue,linkcolor=blue,urlcolor=blue,pdfstartview=FitH,bookmarksopen=true}

\usepackage{color}
\definecolor{zzz}{rgb}{0.9,0.0,0.4}

\begin{document}
\title{Multiple single-photon generations in three-level atoms coupled to cavity with non-Markovian effects}

\author{H. Z. Shen,$^{1,2,}$\footnote{ \textcolor{zzz}{Corresponding author: shenhz458@nenu.edu.cn }}, Y. Chen,$^{1}$  T. Z. Luan,$^{1}$ and X. X. Yi$^{1,2,}$\footnote{ \textcolor{zzz}{Corresponding author: yixx@nenu.edu.cn }}}
\affiliation{$^1$Center for Quantum Sciences and School of Physics,
Northeast Normal University, Changchun 130024, China\\
$^2$Center for Advanced Optoelectronic Functional  Materials
Research, and Key Laboratory for UV Light-Emitting Materials and
Technology of Ministry of Education, Northeast Normal University,
Changchun 130024, China}

\date{\today}

\begin{abstract}
In this paper, we show how to generate the multiple single-photon wavepackets of arbitrary
temporal shape from an optical cavity coupled with $N$ three-level atoms driven by a
driving field in the non-Markovian regime. We derive an exact analytical expression of
the optimal driving field for generating such wavepackets, which depends on two
detunings of the cavity and driving field with respect to the three-level atoms.
The cavity we used consists of two mirrors facing each other, where one is perfect
and the other exists the dissipation (one-sided cavity), which couples with the
corresponding non-Markovian input-output fields. If the first single-photon wavepacket generated by the Markovian system is the same as the non-Markovian case, the Markovian system cannot generate the same multiple single-photon wavepackets as the non-Markovian one when the spectral widths of the other environments taking values different from the spectral width of the first environment, while setting the equal spectral widths for the different environments can generate this. The generated multiple different single-photon wavepackets are not independent of each other, which
satisfy certain relations with non-Markovian spectral parameters. We analyse the
transition from Markovian to non-Markovian regimes and compare the differences
between them, where the cavity interacts simultaneously with the multiple non-Markovian environments.
Finally, we extend the above results to a general non-Markovian quantum
network involving many cavities coupled with driven three-level atoms.
\end{abstract}


\maketitle
\section{Introduction}
Quantum networks composed of local nodes, which are
connected by quantum channels, are essential for quantum
communication and desirable for scalable and distributed
quantum computation \cite{Cirac199959,DiVincenzo200048,Cirac199778,
Hong200776,Zhou200979,vanEnk199778,Clark200391,Chen200776,Nohama200841,Cho200595,Gogyan201898,Peng2000265}.
The stationary qubits in local nodes can be provided by collective atomic excitations
\cite{Schlosshauer2014,Lukin4232,Fleischhauer200265}.
A photon wavepacket is an ideal carrier for a flying qubit, with
either the photon number states or the polarization forming the
qubit.
The controlled production of single photons \cite{Mei200980,Xiao200674} is of fundamental
and practical interest, which denotes the lowest excited quantum
states of the radiation field, and has applications in quantum
information processing \cite{Knill2001409}.
The single-photon generation \cite{Welakuh2017403,Sun200469,Brattke200186,Missori200740}
by a coupled atom cavity \cite{Lai2008281,Xiao200810,vanEnk20043,
Hu201591,Li2010374,Deng2010,Xu201633,Fong201796,Reiserer201587,Ma200595,Meng201919,Ye200673,Wang201999,Schmidt2013525,Schmid201184,Yu2011284,Cho200572} system
has been demonstrated in the Markovian case. After that, several similar schemes have been put forward \cite{Martin2010465,Specht2011473,Dilley201285,Wang201795,Shen201388,Agarwal201693,
Akram201548,Oliveira201817,Liu201059,Borges201694,Thyrrestrup2123130,
Graefe88033842,Manukhova96023851,Borges97013828,Veselkova99013814,Goto99053843,Macha101053406,Cai104053707,Holleczek117023602,Morin123133602,Vitanov89015006,Keller2004431}.
The prototype quantum interface for this purpose was
proposed by Yao et al. \cite{Yao200595}, where the presented Raman process can be made to generate or annihilate \cite{Yao200595,Yao20057,Duan200367,Hong201716,Hong201281}
an arbitrarily shaped single-photon wavepacket by pulse shaping the controlling laser field.

Markovian processes successfully describe many physical phenomena, especially in the field of quantum optics, but they fail when they are applied to more complex system-environment couplings, where memory effects play the dominating roles. Generally speaking, all realistic quantum systems are open due to the unavoidable couplings to environment (of memory or memoryless) \cite{Li201081,Gardiner2000,Franco201327,Caruso201486}.
Considering the non-Markovian \cite{Shen201796,Hong201286,Zhang201387,Breuer2002,xiong201286,
yu199960,Jack199959,Cohen-Tannoudji197710,Zhang200186,Shen201897,Shen201898,Shen201999,Man201592,Hartmann2849,Pellizzari795242,Biswas062303,Turchette053807,Myatt403269,Maniscalco052101,Bayindir2140,Stefanou12127,Xu7389,Lin165330,Chang052105,Tan032102,Longhi063826,xue860523042012,breuer880210022016,breuer1032104012009,laine810621152010,wibmann860621082012,lorenzo880201022013,rivas1050504032010,wolf1011504022008,lu820421032010,chruscinski1121204042014} dynamics of open quantum systems is essential in quantum information technology.
In particular, a notion of memory for quantum processes has been introduced, which
can be physically interpreted in terms of information flow
between the open system and its environments.
So far, several scenarios have been recognized under which the non-Markovian dynamics can happen, for example, strong system-environment coupling, structured reservoirs,
low temperatures, and initial system-environment correlations
\cite{Hong200878,Xiao200673,Wang200775,Hong201569,Laine201092,Dajka201082,Smirne201082,Man2012376,Li201183}.
Generally, people focus on the quantum system coupled to a single environment, which
has been investigated theoretically \cite{Vega201789,Verstraete20095,Kastoryano2013110,Hou201591,Shen201693,
Dariusz2010104,Xu2010104,Chin2012109,Deffner2013111,Leggett198759,Weiss2008} and experimentally
\cite{blacher20156,Liu20117,Ulrich2012108,Xu20101,Xu201082,Madsen2011106,Orieux20155,Smirne201184}.
However, in the real world, there might be a situation of many environments
coupling to a system simultaneously
\cite{Hanson2008320,Hanson200779,Pla2012489,Agarwal201592,Li2011284,Man201592,
Chan201489,Apollaro201490,Yoshie2004432,Vahala2003424,Waks200696,Garnier200775,Propp201927}.

The above two considerations motivate us to explore the generations of
multiple complex single-photon wavepackets from an optical cavity coupled to the driven three-level atoms with non-Markovian input-output fields.

In this paper, we present a scheme of generating the multiple complex
single-photon wavepackets from the cavity coupled with $N$ driven
three-level atoms in the non-Markovian regime. To generate such
wavepackets, we derive an analytical solution of the optimal
driving field, which is affected by two detunings of the
cavity and driving field with respect to the three-level atoms.
When the first single-photon wavepackets generated by the Markovian and non-Markovian systems are equal, the same multiple single-photon wavepackets in the  non-Markovian regime cannot be generated by a Markovian system if all the other spectral widths do not equal the first one, while setting the equal spectral widths for the different environments can generate this.
Moreover, we study the non-Markovian dynamics of
generating the multiple output single-photon wavepackets
from one side of the cavity coupled simultaneously with
multiple (or two) identical and different non-Markovian
input-output fields, which exist certain correlations
related to non-Markovian spectral parameters. Finally,
the above results are extended to a non-Markovian input-output
quantum network consisting of many cavities containing driven
three-level atoms.

Our paper is outlined as follows. In Sec.~{\rm II}, we first illustrate a model to describe the driven three-level atoms coupled to a cavity, which interacts simultaneously with the multiple non-Markovian input-output fields.~In Sec.~{\rm III}, we present the exact solutions of the optimal driving field for generating the multiple
single-photon wavepackets. In Sec.~{\rm IV}, we compare the difference of a single-photon generation with and without the Markovian approximations. In Sec.~{\rm V}, we study the multiple single-photon generations, where the cavity simultaneously interacts with the multiple non-Markovian input-output fields by taking the equal and non-equal values for the  spectral widths of the different environments in the Markovian and non-Markovian systems. Sec.~{\rm VI} is devoted to the discussion of the non-Markovian quantum input-output network with many driven atom-cavity systems. In Sec.~{\rm VII}, we summarize the paper.
\section{MODEL and exact non-Markovian dynamics}
\begin{figure}[h]
\centering
\includegraphics[scale=0.40]{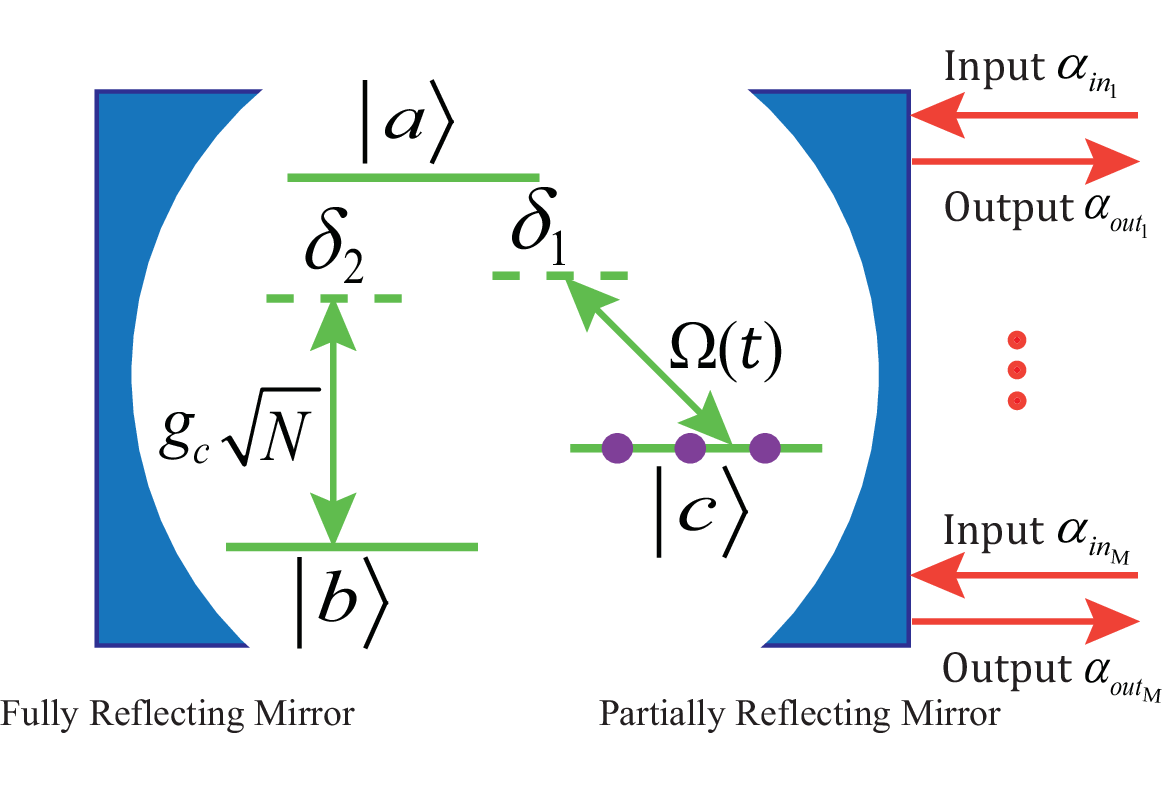}
\caption{Multiple complex single-photon generations in three-level atoms coupled to a cavity with non-Markovian effects. The system is composed of one-sided cavity coupled to $N$ three-level $\Lambda $ atoms, where the cavity interacts simultaneously with $M$ non-Markovian input-output fields. Every atom has three levels, i.e., the ground
state hyperfine levels $\left| b \right\rangle $, $\left| c \right\rangle $ \cite{McKeever2003425,McKeever2004303}, and the excited state $\left| a \right\rangle $, as shown in Fig.~\ref{MODEL}. Three-level atoms interact with cavity and are driven by classical field with the coupling strengths ${g_{c}}\sqrt N $ and driving field $\Omega (t)$, respectively. The purple dots denote schematically that the atomic population is concentrated in the state $\left| c \right\rangle $. The classical field and cavity are detuned from the atoms denoted by ${\delta _1}$ and ${\delta _2}$,
respectively.} \label{MODEL}
\end{figure}

The proposed scheme for the non-Markovian multiple complex single-photon
generations is depicted in Fig.~\ref{MODEL}, where a Fabry-P{\'e}rot cavity
couples to $N$ identical three-level $\Lambda $-type atoms in the basis of
collective states. We now discuss how to generate the  shapes of the
specified single-photon wavepackets if there are no incoming photons. The single-photon
pulse shapes, provided they are smooth enough, can be arbitrarily specified.
The total system is described by the
Hamiltonian $\hat H = \hat H_S + {\hat H}_B + \hat V$ in
the rotating frame (setting $\hbar  \equiv 1$):
\begin{equation}
\begin{aligned}
{\hat H_S} =& \sum\limits_{m = 1}^N {\{ [\Omega } (t){e^{i{\delta _1}t}}\hat \sigma _{ac}^{(m)} + {g_c}\hat \sigma _{ab}^{(m)}\hat a{e^{i{\delta _2}t}} + H.c.]\\
&- i\gamma '\hat \sigma _{aa}^{(m)}\},\\
{{\hat H}_B} =& \sum\limits_{j = 1}^M {\int {d{\omega _j}} } {\Omega _{{\omega _j}}}\hat b_j^\dag ({\omega _j}){{\hat b}_j}({\omega _j}),\\
\hat V =& i\sum\limits_{j = 1}^M {\int {d{\omega _j}[{v _j}} } ({\omega _j})\hat a\hat b_j^\dag ({\omega _j}) - H.c.],
\label{Htotal}
\end{aligned}
\end{equation}
where $\hat \sigma _{\mu ,\nu }^{(m)} = {| \mu  \rangle _{mm}}\negmedspace\langle \nu  |$ $(\mu ,\nu  = a,b,c)$ is the flip operator of the $m$th atom between states $| \mu  \rangle$ and $\langle \nu  |$. $H.c.$ stands for the Hermitian conjugate.
${{\hat b}_j}({\omega _j})$ $(\hat b_j^\dag ({\omega _j}))$ is the annihilation (creation) operator for the frequency
$\omega _j$ in the $j$th continuum field (it can also be called $j$th environment). ${\hat a}$ (${{\hat a}^\dag }$) denotes the annihilation (creation) operator of the
cavity. The interaction between the cavity and continuum field is described by Hamiltonian $\hat V$ with the strength ${v _j}({\omega _j})$ \cite{Gardiner198531,Gardiner2000,Zhang201387,Walls2008}, where $[ {{{\hat b}_j}({\omega _j}),\hat b_j^\dag ({{\omega^{\prime}_j}})} ] = \delta ( {{\omega _j} - {\omega^{\prime}_j}})$
 and $[ {\hat a,{\hat a^\dag }} ] = 1$. The derivation of Eq.~(\ref{Htotal}) can be found in Appendix A.
In view of the symmetry of the couplings, it is convenient to introduce collective atomic operators ${{\tilde { \sigma}} _{ab}} = \sum\nolimits_{m = 1}^N {{\hat \sigma} _{ab}^{(m)}} $, ${{\tilde { \sigma}} _{ac}} = \sum\nolimits_{m = 1}^N {{\hat \sigma} _{ac}^{(m)}} $, and ${{\tilde { \sigma}} _{aa}} = \sum\nolimits_{m = 1}^N {{\hat \sigma} _{aa}^{(m)}} $.
When all atoms are prepared initially in level $| b \rangle $, the only states coupled
by the interaction are totally symmetric Dicke-like states \cite{Rajapakse200980,Xu201388}
\begin{equation}
\begin{aligned}
&\left| b \right\rangle  \equiv \left| {{b_1} \ldots {b_m} \ldots {b_N}}\right\rangle, \\
&\left| a \right\rangle  \equiv {\frac{1}{{\sqrt N }}} \sum\limits_{m = 1}^N  | {{a_m}\prod\limits_{ \otimes k = 1}^N {{b_{k \ne m}}} } \rangle  ,\\
&\left| c \right\rangle  \equiv {\frac{1}{{\sqrt N }}} \sum\limits_{m = 1}^N  | {{c_m}\prod\limits_{ \otimes k = 1}^N {{b_{k \ne m}}} } \rangle ,
\label{numberstate}
\end{aligned}
\end{equation}
where the introduced factor ${1/{\sqrt N }}$ in front of Eq.~(\ref{numberstate}) meets the state normalization requirements, i.e., $\langle {a|a} \rangle  =\langle {c|c} \rangle = 1$. The $| b \rangle  \to | a \rangle $
transition is coupled to the cavity with the strength ${g_{c}}$, which is assumed to be equal for all atoms.
The detuning ${\delta _2}$ is defined as ${\delta _2} = {\omega _a} - {\omega _b} - {\omega _{cav}}$
(${\omega _{cav}}$ is the center frequency of the cavity).
The $| c \rangle $$\to | a \rangle $ transition is
coupled by the time-dependent driving field $\Omega ( t )$ (i.e., the Rabi frequency of the driving field) \cite{Scully1997} with the detuning ${\delta _1} = {\omega _a} - {\omega _c} - {\omega _L}$, where ${\omega _L}$
is the frequency of the classical field, while
${\omega _b}$, ${\omega _c}$ and ${\omega _a}$ denote respectively the eigenfrequencies of states $| b \rangle $, $| c \rangle $ and $| a \rangle $.
$\gamma '$ is the atomic spontaneous emission rate. ${\Omega _{{\omega _j}}} = {\omega _j} - {\omega _{cav}}$ denotes the detuning
of the $\omega_j$-mode of the continuum fields from the center frequency of the cavity.
The atom-cavity interacts with input-output fields by bases $| {b,1,0} \rangle $,
{$| {c,0,0} \rangle $, $| {a,0,0} \rangle $ and
$| {b,0,{1_{{\omega _j}}}} \rangle $},
respectively, where $|s,n,0\rangle {\rm{ = }}|s\rangle  \otimes |n\rangle  \otimes |0\rangle $ and $|s,n,{1_{{\omega _j}}}\rangle {\rm{ = }}|s\rangle  \otimes |n\rangle  \otimes |{1_{{\omega _j}}}\rangle  $ ($s = a, b, c$, and $n$ is the number of photons in the cavity).
$| 0 \rangle  = | {\cdots{0_{_1}},{0_{_2}}, \cdots {0_{_M}\cdots}} \rangle $
corresponds to the continuum fields at its vacuum state. ${| 1_{{\omega _j}}} \rangle  = | {\cdots{0_{_1}},{0_{_2}}, \cdots {1_{{\omega _j}}} \cdots {0_M}\cdots} \rangle$
 denotes the one-photon Fock state of the continuum fields with frequency ${\omega _j}$.
The state of the total system can be
expressed in the compact form:
\begin{equation}
\begin{aligned}
| {{\Psi }(t)} \rangle  = &{\beta _b}(t)| {b,1,0} \rangle  + {\beta _c}(t)| {c,0,0} \rangle  + {\beta _a}(t)| {a,0,0} \rangle \\
&+ \sum\limits_{j = 1}^M {\int {d{\omega _j}} } {\alpha _{{\omega _j}}}(t)| {b,0,{1_{{\omega _j}}}} \rangle.
\label{totalstates}
\end{aligned}
\end{equation}
With these relations ${{\tilde \sigma }_{ba}}| a \rangle  = \sqrt N | b \rangle$, ${{\tilde \sigma }_{ab}}| b \rangle  = \sqrt N | a \rangle $, ${{\tilde \sigma }_{ca}}| a \rangle  = | c \rangle $, and ${{\tilde \sigma }_{ac}}| c \rangle  = | a \rangle $, substituting Eqs.~(\ref{Htotal}) and~(\ref{totalstates}) into Schr{\"o}dinger equation $i|\dot { \Psi}  \left( t \right)\rangle  = \hat H|\Psi  \left( t \right)\rangle $, we obtain a set of the differential equations for the probability amplitudes
\begin{small}
\begin{eqnarray}
{{\dot \beta }_b}(t) &=& - i{g_c}\sqrt N {e^{ - i{\delta _2}t}}{\beta _a}(t) - \sum\limits_{j = 1}^M {\int_{ - \infty }^{ + \infty } {d{\omega _j}v_j^*({\omega _j})} } \nonumber\\
 &&\times {\alpha _{\omega j}}(t),\label{generalEq1}\\
{{\dot \beta }_c}(t) &=& - i{\Omega ^*}(t){e^{ - i{\delta _1}t}}{\beta _a}(t),\label{generalEq2}\\
{{\dot \beta }_a}(t) &=& - i{g_c}\sqrt N {e^{i{\delta _2}t}}{\beta _b}(t) - i\Omega (t){e^{i{\delta _1}t}}{\beta _c}(t) \nonumber\\
 &&- \gamma '{\beta _a}(t),\label{generalEq3}\\
{{\dot \alpha }_{{\omega _j}}}(t) &=&  - i{\Omega _{{\omega _j}}}{\alpha _{{\omega _j}}}(t) + {v_j}({\omega _j}){\beta _b}(t).\label{general}
\end{eqnarray}
\end{small}
Integrating Eq.~(\ref{general}), we obtain
\begin{equation}
\begin{aligned}
{\alpha _{{\omega _j}}}(t) = &{e^{ - i{\Omega _{{\omega _j}}}t}}{\alpha _{{\omega _j}}}({0}) + {v _j}({\omega _j})\int_{{0}}^t {d\tau {\beta _b}(\tau )} \\
& {e^{ - i{\Omega _{{\omega _j}}}(t - \tau )}},
\label{awjt0}
\end{aligned}
\end{equation}
or
\begin{equation}
\begin{aligned}
{\alpha _{{\omega _j}}}(t) = &{e^{ - i{\Omega _{{\omega _j}}}(t - {t_1})}}{\alpha _{{\omega _j}}}({t_1}) - {v _j}({\omega _j})\int_t^{{t_1}} {d\tau {\beta _b}(\tau )} \\
& {e^{ - i{\Omega _{{\omega _j}}}(t - \tau )}},
\label{awjt1}
\end{aligned}
\end{equation}
where ${t_1} \ge t$.
Substituting Eq.~(\ref{awjt0}) into Eq.~(\ref{generalEq1}), we get the non-Markovian integro-differential equations for the probability amplitudes
\begin{small}
\begin{equation}
\begin{aligned}
{{\dot \beta }_b}(t) =& - i{g_c}\sqrt N {e^{ - i{\delta _2}t}}{\beta _a}(t) - \sum\limits_{j = 1}^M {\int_0^t {d\tau {F_j}(t - \tau )} }   {\beta _b}(\tau ) \\
 & + \sum\limits_{j = 1}^M {\int {d\tau k_j^*(\tau  - t){\alpha _{i{n_j}}}(\tau )} },\\
{{\dot \beta }_c}(t) =& - i{\Omega ^*}(t){e^{ - i{\delta _1}t}}{\beta _a}(t),\\
{{\dot \beta }_a}(t) =& - i{g_c}\sqrt N {e^{i{\delta _2}t}}{\beta _b}(t) - \gamma '{\beta _a}(t) - i\Omega (t){e^{i{\delta _1}t}}{\beta _c}(t).
\label{finallyEq1}
\end{aligned}
\end{equation}
\end{small}
The input fields ${\alpha _{i{n_j}}}(t)$ in the non-Markovian regime related to the output fields ${\alpha _{ou{t_j}}}(t)$ by the multiple input-output relations are derived as follows
\begin{eqnarray}
{\alpha _{i{n_j}}}(t) + {\alpha _{ou{t_j}}}(t)= \int_{{0}}^t {d\tau {k_j}(t - \tau ){\beta _b}(\tau )}
\label{finallyEq4},
\end{eqnarray}
where
\begin{eqnarray}
{\alpha _{i{n_j}}}(t) =  - \frac{1}{{\sqrt {2\pi } }}\int_{ - \infty }^{ + \infty } {d{\omega _j}} {\alpha _{{\omega _j}}}(0){e^{ - i{\Omega _{{\omega _j}}}t}}
\label{alphainj},
\end{eqnarray}
and
\begin{eqnarray}
{\alpha _{ou{t_j}}}(t) = \frac{1}{{\sqrt {2\pi } }}\int_{ - \infty }^{ + \infty } {d{\omega _j}} {\alpha _{{\omega _j}}}({t_1}){e^{ - i{\Omega _{{\omega _j}}}(t - {t_1})}}{\rm{ }}.
\label{alphaoutj}
\end{eqnarray}
The details of the derivation in Eqs.~(\ref{finallyEq1}) and (\ref{finallyEq4}) can be found in Appendices B and C, respectively.
The response function and correlation function can be respectively written as
\begin{equation}
\begin{aligned}
{k_j}(t) = \frac{1}{{\sqrt {2\pi } }}\int_{ - \infty }^\infty {d{\omega _j}} e{}^{ - i{\Omega _{\omega_j}}t}{v _j}({\omega _j}),
\label{kj}
\end{aligned}
\end{equation}
and
 \begin{equation}
\begin{aligned}
{F_j}(t - \tau ) = \int_{ - \infty }^{ + \infty } {d{\omega _j}} {J_j}({\omega _j}){e^{ - i{\Omega _{{\omega _j}}}(t - \tau )}},
\label{f}
\end{aligned}
\end{equation}
where ${J_j}({\omega _j}) = {| {{v _j}({\omega _j})} |^2}$ represents the
spectral density. We assume ${v_j}({\omega _j}) = {\lambda _j}\sqrt {{\gamma _j}/2\pi } /[{\lambda _j} - i({\omega _j} - {\omega _{cav}})]$ and
\begin{equation}
\begin{aligned}
{J_j}({\omega _j}) = \frac{{{\gamma _j}}}{{2\pi }}\frac{{\lambda _j^2}}{{\lambda _j^2 + {{({\omega _j} - {\omega _{cav}})}^2}}},
\label{Jjtt}
\end{aligned}
\end{equation}
where $\gamma _j $ and $\lambda _j$ denote the decay rate and spectral width of the $j$th environment, respectively. In the non-Markovian regime, we have
${k_j}(t) = {\lambda _j}\sqrt {{\gamma _j}} u(t){e^{ - {\lambda _j}t}}$ and
${F_j}(t) = \frac{1}{2}{\lambda _j}{\gamma _j}{e^{ - {\lambda _j}\left| t \right|}}$, where $u (t)$ is the unit step function, i.e., $u(t) = 1$ for $t \ge 0$, otherwise $u(t) = 0$. Under the Markovian approximation, the spectral density $J_j({\omega _j})\to{\gamma _j}/2\pi$ and coupling strength ${v_j}({\omega _j})\to\sqrt {{\gamma _j}/2\pi }$  lead to
\begin{equation}
\begin{aligned}
{k_j}(t) &= \sqrt {{\gamma _j}} \delta (t),\\
{F_j}(t) &= {\gamma _j}\delta (t).
\label{markovkjfj}
\end{aligned}
\end{equation}
Substituting Eq.~(\ref{markovkjfj}) into Eqs.~(\ref{finallyEq1}) and~(\ref{finallyEq4}), we
obtain
\begin{small}
\begin{equation}
\begin{aligned}
&{{\dot \beta }_b}(t) =  - i{g_c}\sqrt N {e^{ - i{\delta _2}t}}{\beta _a}(t) + \sum\limits_{j = 1}^M {\sqrt {{\gamma _j}} {\alpha _{i{n_j}}}(t)}  - \sum\limits_{j = 1}^M {\frac{1}{2}{\gamma _j}{\beta _b}(t)}, \\
&{{\dot \beta }_c}(t) =  - i{\Omega ^*}(t){e^{ - i{\delta _1}t}}{\beta _a}(t),\\
&{{\dot \beta }_a}(t) =  - i{g_c}\sqrt N {e^{i{\delta _2}t}}{\beta _b}(t) - \gamma '{\beta _a}(t) - i\Omega (t){e^{i{\delta _1}t}}{\beta _c}(t),\\
&{\alpha _{in_j}}(t) + {\alpha _{out_j}}(t) = \sqrt {{\gamma _j}} {\beta _b}(t).
\label{lastnonmarkov}
\end{aligned}
\end{equation}
\end{small}
We show that from Eq.~(\ref{finallyEq1}) the probability
amplitudes and driving field are expressed in terms of ${\alpha _{in_j}}(t)$
and ${\alpha _{out_j}}(t)$, which can be arbitrarily specified and
generated on demand by meeting
normalization condition of Eq.~(\ref{totalstates}).
\section{non-Markovian multiple single-photon generations and optimal driving field}
In the following, we will discuss in more detail that
the system can generate the multiple complex single-photon wavepackets.
If initially the three-level system is entirely in state
$\left| {c,0,0} \right\rangle $, this mapping operation can function as the deterministic
generations of the single-photon wavepackets with any desired
pulse shape ${{\alpha _{out_j}}(t)}$.
The initial conditions for the above scheme take ${\alpha _{in_j}}(t)=0$,
${\beta _b}(0) = 0$, ${\beta _c}(0) = 1$, and ${\beta _a}(0) = 0$, together with Eqs.~(\ref{finallyEq1}) and (\ref{finallyEq4}), which lead to
\begin{eqnarray}
{\beta _b}(t) &=& \frac{{{{\dot \alpha }_{out_j}}(t) + {\lambda _j}{\alpha _{out_j}}(t)}}{{{\lambda _j}\sqrt {{\gamma _j}} }},
\label{Bbt}\\
{{\tilde \beta }_a}(t) &=& \frac{{[ - {{\dot \beta }_b}(t) - \sum\limits_{j = 1}^M {\int_0^t {d\tau {F_j}(t - \tau ){\beta _b}(\tau )]} } }}{{{g_c}\sqrt N }},
\label{Bat}
\end{eqnarray}
where we have defined ${{\tilde \beta }_a}(t) = i{e^{ - i{\delta _2}t}}{\beta _a}(t)$.
In order not to lose generality, we assume that the target pulse shapes of the output single-photon wavepackets are
specified to be complex functions in time with the interaction picture so that ${\beta _b}(t)$ and ${{\tilde \beta }_a}(t)$ are also complex ones in this case.
According to Eq.~(\ref{finallyEq1}), we get
\begin{eqnarray}
\Omega (t){{\tilde \beta }_c}(t) &=& {\dot{\tilde \beta }_a}(t) + i{\delta _2}{{\tilde \beta }_a}(t) - {g_{c}}\sqrt N {\beta _b}(t)\nonumber\\
&& + {\gamma '}{{\tilde \beta }_a}(t),
\label{omegabc}\\
{\Omega ^*}(t){{\tilde \beta }_a}(t) &=&  - i\tilde \delta {{\tilde \beta }_c}(t) - {\dot{\tilde \beta }_c}(t),
\label{omegaxba}
\end{eqnarray}
where ${\tilde \beta _c}(t) = {e^{ - i\tilde \delta t}}{\beta _c}(t)$ and $\tilde \delta = {\delta _2} - {\delta _1}$. Based on Eqs.~(\ref{omegabc}) and (\ref{omegaxba}), we obtain the population of the atom in the state $| c \rangle $ with non-Markovian effects
\begin{equation}
\begin{aligned}
{\rho _c}(t) =& 1 - {| {{{\tilde \beta }_a}(t)} |^2}\\
 &+ \int_0^t {dt'} {\rm{\{ }}2{g_c}\sqrt N {\rm{Re}}[{{\tilde \beta }_a}(t')\beta _b^*(t')] - 2\gamma '|{{\tilde \beta }_a}(t'){|^2}{\rm{\} }},
\label{A}
\end{aligned}
\end{equation}
where ${\rho _c}(t) = {| {{{\tilde \beta }_c}(t)} |^2}$, which does not depend on the detunings ${\delta _1}$ and ${\delta _2}$. The strategy is to design the  driving field $\Omega (t)$ from Eqs.~(\ref{omegabc}) and (\ref{omegaxba}) as follows
\begin{equation}
\begin{aligned}
\Omega (t) = \chi (t)[{{\dot {\tilde \beta }_a}}(t) + i{\delta _2}{{\tilde \beta }_a}(t) - {g_c}\sqrt N {\beta _b}(t) + \gamma '{{\tilde \beta }_a}(t)],
\label{omegadelta}
\end{aligned}
\end{equation}
to generate the multiple single-photon wavepackets with non-Markovian effects, where $\chi (t) = \exp \int_0^t {dt'} \{ [ i \tilde \delta {\rho _c}(t') + {{\tilde \beta }_a}(t'){\dot {\tilde {\beta _a^*}}} (t') -i {\delta _2}{| {{{\tilde \beta }_a}(t')} |^2} - {g_c}\sqrt N {{\tilde \beta }_a}(t')\beta _b^ * (t') + \gamma '{| {{{\tilde \beta }_a}(t')} |^2}]/{\rho _c}(t')\}$. In this case, we find that the optimal driving field $\Omega (t)$ depends on  two detunings ${\delta _1}$ and ${\delta _2}$ for the generations of the multiple complex single-photon wavepackets (it will be exhibited in the ending of Sec.~IV), which completely differs from the previous proposals \cite{Hong200776,Keller2004431,Goto99053843,Manukhova96023851,Chen200776,Cirac199778,Yao200595,Yao20057,Duan200367,Hong201716,Hong201281}.

In particular, assuming that the output single-photon wavepackets are real functions with the time in the interaction picture so that ${\beta _b}(t)$ and ${{\tilde \beta }_a}(t)$ also are real ones, we obtain the exact analytical expression
for the optimal driving field to generate the desired output   single-photon wavepackets as $
\Omega (t) = P(t) + iQ(t),$
whose argument and modulus  are respectively expressed by
\begin{eqnarray}
&&\arg [\Omega (t)] = \arctan [\frac{{Q(t)}}{{P(t)}}],
\label{omegaaaa}\\
&&\left| {\Omega (t)} \right|  = \sqrt {\frac{{{[{\dot{\tilde \beta }_a}(t) - {g_{c}}\sqrt N {\beta _b}(t) + \gamma '{{\tilde \beta }_a}(t)]^2} + \delta _2^2{\tilde \beta ^2}_a(t)}}{{{\rho _c}(t)}}},\nonumber \\
\label{omegamodulus}
\end{eqnarray}
where $
P(t) = [{\dot{\tilde \beta }_a}(t)\cos \alpha (t) - {g_{c}}\sqrt N {\beta _b}(t)\cos \alpha (t)+ {\gamma '}{{\tilde \beta }_a}(t)\cos \alpha (t) + {\delta _2}{{\tilde \beta }_a}(t)\sin \alpha (t)]/\sqrt {\rho _{c}(t)}$ and $Q(t) = [{\delta _2}{{\tilde \beta }_a}(t)\cos \alpha (t) + {g_{c}}\sqrt N {\beta _b}(t)\sin \alpha (t) - {\dot{\tilde \beta }_a}(t)\sin \alpha (t) - {\gamma '}{{\tilde \beta }_a}(t)\sin \alpha (t)]/\sqrt {\rho _{c}(t)}$
with $\alpha (t) =  - \tilde \delta  \cdot t + {\delta _2}\int_0^t {\tilde \beta _a^2(t')/{\rho _c}(t')} dt'.$
Eqs.~(\ref{omegaaaa}) and (\ref{omegamodulus}) tell us that the modulus of the optimal driving field $\Omega(t)$
only has a bearing on the detuning ${{\delta _2}}$, while the detunings ${\delta _1}$ and ${\delta _2}$ produce the influences on the argument $\arg [\Omega (t)]$. Based on these, we show that
the scheme under study for any given multiple single-photon wavepackets
requires the driving field depending on two detunings (i.e., the detunings $\delta_1$
and $\delta_2$ of the cavity and driving field with respect to the three-level atoms) and non-Markovian effects, which completely differs from those of
three-level system shown in Refs.\cite{Hong200776,Keller2004431,Goto99053843,Manukhova96023851,Chen200776,Cirac199778,Yao200595,Yao20057,Duan200367,Hong201716,Hong201281}, where these schemes mainly focused on the resonance case ($\delta_1=\delta_2\equiv0$) and Markovian approximation.

Under the Markovian approximation with Eq.~(\ref{lastnonmarkov}) (we denote the quantities in the Markovian case by introducing a subscript $f$ to them), subjected to the initial conditions ${\alpha _{in_j}}(t)=0$, ${\beta _{bf}}(0) = 0$, ${\beta _{cf}}(0) = 1$, and ${\beta _{af}}(0) = 0$, we can obtain
\begin{small}
\begin{equation}
\begin{aligned}
{\beta _{bf}}(t) =& \frac{{{\alpha _{ou{t_{jf}}}}(t)}}{{\sqrt {{\gamma _j}} }},\\
{{\tilde \beta }_{af}}(t){\rm{ }} =& \frac{{[ - {{\dot \beta }_{bf}}(t) - \sum\limits_{j = 1}^M {\frac{1}{2}{\gamma _j}{\beta _{bf}}(\tau )} ]}}{{{g_c}\sqrt N }},\\
{\rho _{cf}}(t) =& 1 - |{{\tilde \beta }_{af}}(t){|^2}\\
&+ \int_0^t {dt'} \{ 2{g_c}\sqrt N {\rm{Re}}[{{\tilde \beta }_{af}}(t')\beta _{bf}^*(t')] - 2\gamma '|{{\tilde \beta }_{af}}(t'){|^2}\},\\
{\Omega _f}(t) =& {P_f}(t) + i{Q_f}(t),
\label{lastmarkov}
\end{aligned}
\end{equation}
\end{small}
with
\begin{small}
\begin{equation}
\begin{aligned}
{P_f}(t) =& [{\dot{\tilde \beta } _{af}}(t)\cos {\alpha _f}(t) - {g_c}\sqrt N {\beta _{bf}}(t)\cos {\alpha _f}(t)\\
& + \gamma '{\tilde \beta _{af}}(t)\cos {\alpha _f}(t) + {\delta _2}{\tilde \beta _{af}}(t)\sin {\alpha _f}(t)]/\sqrt {{\rho _{cf}}(t)},\\
{Q_f}(t) =& [{\delta _2}{\tilde \beta _{af}}(t)\cos {\alpha _f}(t) + {g_c}\sqrt N {\beta _{bf}}(t)\sin {\alpha _f}(t)\\
& - {\dot{\tilde \beta } _{af}}(t)\sin {\alpha _f}(t) - \gamma '{\tilde \beta _{af}}(t)\sin {\alpha _f}(t)]/\sqrt {{\rho _{cf}}(t)},\\
{\alpha _f}(t) =&  - \tilde \delta  \cdot t + {\delta _2}\int_0^t {\frac{{\tilde \beta _{af}^2(t')}}{{{\rho _{cf}}(t')}}} dt',
\end{aligned}
\end{equation}
\end{small}
where we have defined ${{\tilde \beta }_{af}}(t) = i{e^{ - i{\delta _2}t}}{\beta _{af}}(t)$, ${\tilde \beta _{cf}}(t) = {e^{ - i\tilde \delta t}}{\beta _{cf}}(t)$, and ${\rho _{cf}}(t) = {| {{{\tilde \beta }_{cf}}(t)} |^2}$.

\section{NUMERICAL INVESTIGATION OF MARKOVIAN AND NON-Markovian CASEs}
\begin{figure}[h]
\centering
\includegraphics[height=3.9cm,width=8.1cm]{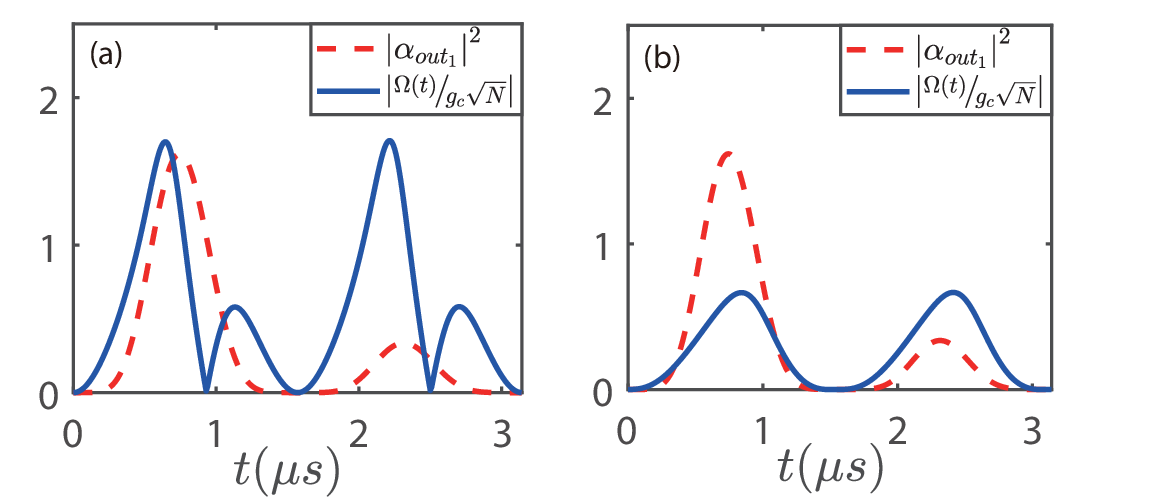}
\caption {(Color online) Generation of the single-photon pulse for the target shape wavepacket ${{\alpha _{ou{t_1}}}(t)}$. The parameters chosen are ${\gamma '} = 6\pi $ MHz, ${g_{c}} = 30\pi$ MHz \cite{McKeever2003425,McKeever2004303}. The other parameters take ${\gamma _1} = 10$ MHz, ${\delta _1} ={\delta _2} = 0$, $N = 40$, ${B_1}=2$ MHz, and ${\Gamma_1} = 0.5$ MHz.
The system initially is prepared in state $\left| {c,0,0} \right\rangle $. For comparison,
$| {\Omega (t)/({g_{c}}\sqrt N )} |$(blue-solid lines) and ${\left| {{\alpha _{ou{t_1}}}(t)} \right|^2}$ (red-dashed lines) are plotted in (a) and (b), where ${{\alpha _{ou{t_1}}}(t)}$ and $\Omega (t)$ are respectively given by Eqs.~(\ref{aoujtttttt}) and (\ref{OMEGA}).
Note that in this case, we choose ${\lambda _1} = 2.31$ MHz (corresponding to the non-Markovian regime) for (a), ${\lambda _1} =30$ MHz (corresponding to the Markovian approximation) for (b).}
\label{ALPHAOUT2}
\end{figure}

\begin{figure}[h]
\centering
\includegraphics[height=6.1cm,width=7.5cm]{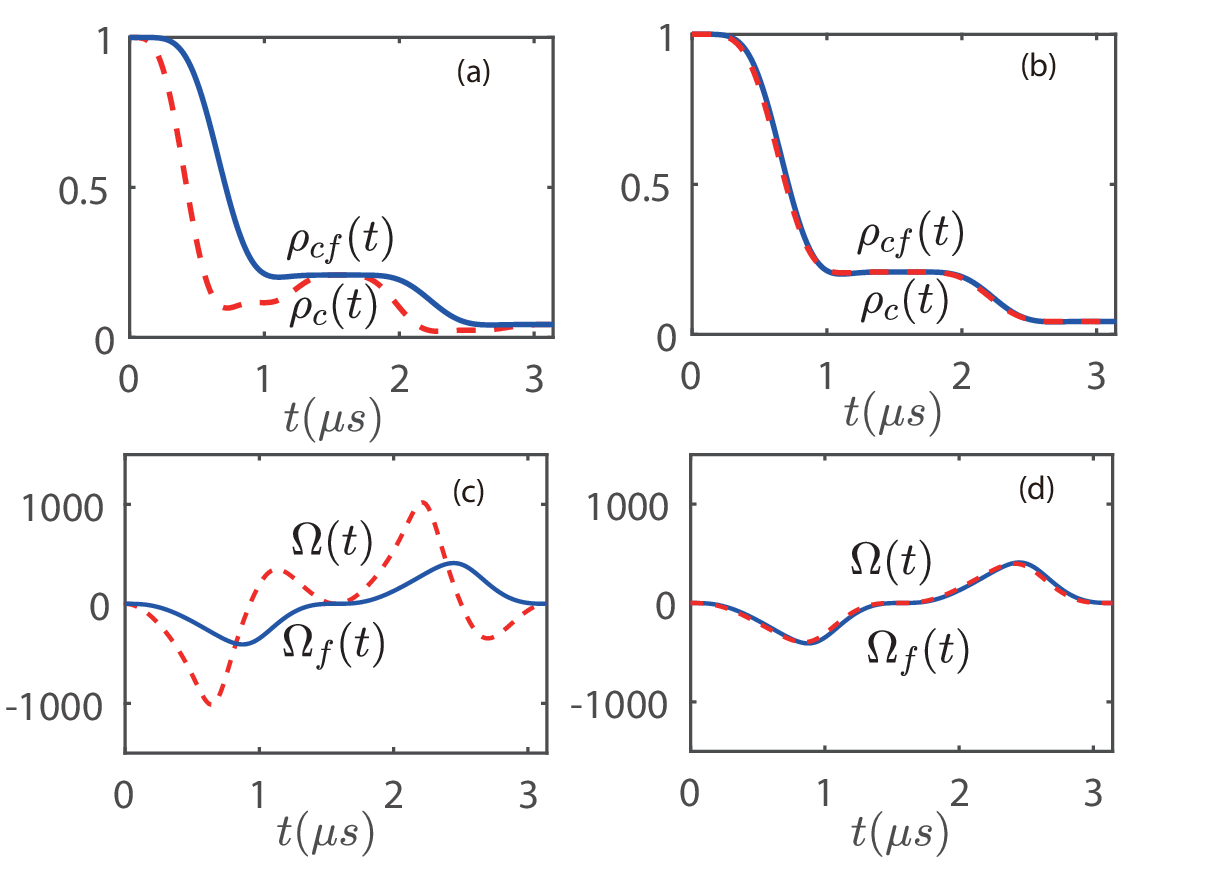}
\caption{(Color online) The populations ${\rho _{cf}}(t)$ and ${\rho _c}(t)$ of the state $\left| c \right\rangle $ with and without the Markovian approximations are plotted in blue-solid and red-dashed lines in (a) and (b), respectively, where ${\rho _c}(t)$ and ${\rho _{cf}}(t)$ are given by Eqs.~(\ref{A}) and~(\ref{lastmarkov}), respectively. The real optimal driving field designed with and without the
Markovian approximations (i.e., ${\Omega _f}(t)$ and $\Omega (t)$)
are plotted in blue-solid and red-dashed lines in (c) and (d),
respectively, where $\Omega (t)$ and $\Omega_ f (t)$
are determined by Eqs.~(\ref{OMEGA}) and~(\ref{lastmarkov}).
The parameters chosen are ${\gamma _1} = 10$ MHz,
${\gamma '} = 6\pi$ MHz, ${g_{c}} = 30\pi$ MHz, ${\delta _1}={\delta _2} = 0$,
$N = 40$, ${B_1}=2$ MHz, and ${\Gamma_1} = 0.5$ MHz, moreover,
${\lambda _1} = 2.31$ MHz for (a) and (c), ${\lambda _1} =30$ MHz for (b) and (d).}
\label{pcfandomega}
\end{figure}

As the memory effect may be helpful in quantum information processing, the non-Markovian dynamics plays important roles in the description of open systems. Among these topics, the system consisting of
$N$ atoms interacting with the multiple environments is of particular interest.
Therefore, we wish to derive the dynamics of the system coupled to the multiple environments. Our aim is to control the production of $M$ output single-photon wavepackets from the cavity by tuning the driving field. In this section, we set $M=1$, i.e., a   single-photon generation in the non-Markovian regime. The output   single-photon wavepacket needs to satisfy ${\beta _b}(0) = [{{\dot \alpha }_{ou{t_j}}}(0) + {\lambda _j}{\alpha _{ou{t_j}}}(0)]/({\lambda _j}\sqrt {{\gamma _j}} ) \equiv 0$ and ${{\dot \beta }_b}(0) = [{{\ddot \alpha }_{ou{t_j}}}(0) + {\lambda _j}{{\dot \alpha }_{ou{t_j}}}(0)]/({\lambda _j}\sqrt {{\gamma _j}} ) \equiv 0$ (originating from Eq.~(\ref{Bbt}) and its derivative in Eq.~(\ref{finallyEq1}) at $t=0$ under the initial conditions ${\alpha _{in_j}}(t)=0$, ${\beta _b}(0) = 0$, ${\beta _c}(0) = 1$, and ${\beta _a}(0) = 0$), i.e., three conditions ${\alpha _{ou{t_1}}}(0) = 0$, ${\dot \alpha _{ou{t_1}}}(0) = 0$, and ${\ddot \alpha _{ou{t_1}}}( 0 ) = 0$, where the output single-photon wavepacket is assumed as a real single-photon wavepacket (meeting the normalization condition)
\begin{equation}
\begin{aligned}
{{\alpha _{ou{t_1}}}(t)}=A_1{e^{ - \Gamma_1 t}}{\sin ^3} B_1 t,
\label{aoujtttttt}
\end{aligned}
\end{equation}
with the system being initially prepared in $\left| {c,0,0} \right\rangle $, corresponding to ${\beta _c}(0) = 1$ and population of the atom in other states being initially
zero. ${A_1} = 2\sqrt {2(36B_1^6{\Gamma _1} + 49B_1^4\Gamma _1^3 + 14B_1^2\Gamma _1^5 + \Gamma _1^7)} /(3\sqrt 5 B_1^3)$ is the normalization coefficient. In addition, assuming ${\delta _1} = {\delta _2} = 0$, leading to $Q(t) = 0$ and $\alpha(t) = 0$, we get
\begin{equation}
\begin{aligned}
\Omega (t) = [{\dot{\tilde \beta }_a}(t) - {g_{c}}\sqrt N {\beta _b}(t)+ \gamma '{{\tilde \beta }_a}(t)]/\sqrt {{\rho _c}(t)}.
\label{OMEGA}
\end{aligned}
\end{equation}
\begin{figure}[h]
\centering
\includegraphics[height=6.4cm,width=8.2cm]{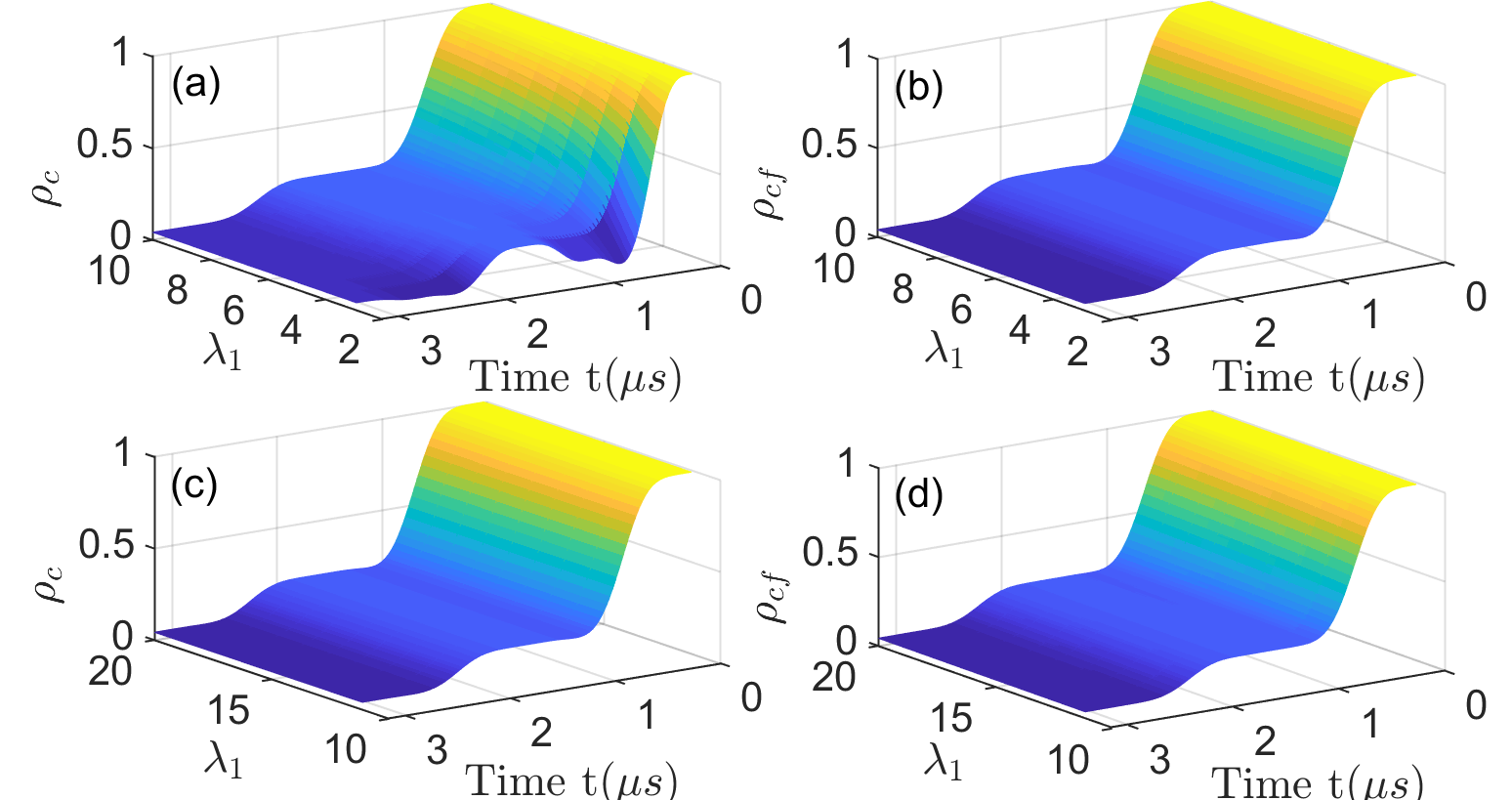}
\caption{(Color online) The evolution of the population of state $\left| c \right\rangle $ as functions of $t$ and $\lambda _1$, where ${\rho _c}(t)$ and ${\rho _{cf}}(t)$ are given by Eqs.~(\ref{A}) and~(\ref{lastmarkov}), respectively. Parameters chosen are ${\gamma _1} = 10$ MHz, ${\gamma '} = 6\pi$ MHz, ${g_{c}} = 30\pi$ MHz, ${\delta _1} ={\delta _2} = 0$, $N = 40$, ${B_1}=2$ MHz, and ${\Gamma_1} = 0.5$ MHz.}
\label{PcPcf}
\end{figure}

\begin{figure}[h]
\centering
\includegraphics[height=9.1cm,width=7.8cm]{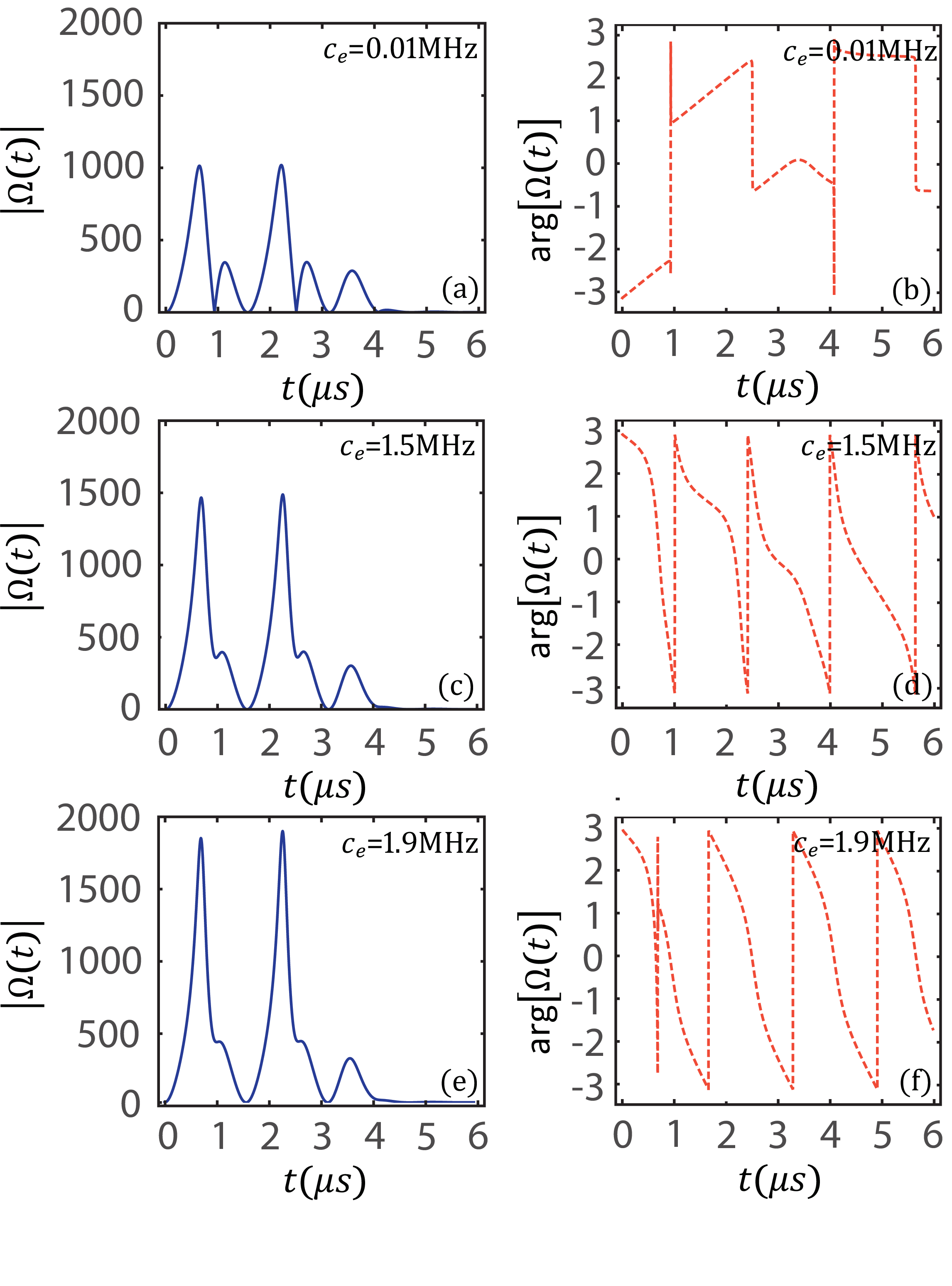}
\caption{(Color online) The influence of the generations of
the complex single-photon wavepacket ${\alpha _{ou{t_1}}}(t) = {A_1}{e^{ - {\Gamma _1}t}}{\sin ^3}({B_1}t){e^{ - i{c_e}t}}$ with ${A_1} = 2\sqrt {2(36B_1^6{\Gamma _1} + 49B_1^4\Gamma _1^3 + 14B_1^2\Gamma _1^5 + \Gamma _1^7)} /(3\sqrt 5 B_1^3)$ on the modulus and argument for the optimal driving field $\Omega (t)$ in the non-Markovian regime, which can be obtained by Eq.~(\ref{omegadelta}). The parameters chosen are ${\lambda _1} = 2.31$ MHz, ${\delta_1} = 1$ MHz, ${\delta_2} = 2$ MHz, ${\gamma _1} = 10$ MHz, ${\gamma '} = 6\pi$ MHz, ${g_{c}} = 30\pi$ MHz, $N = 40$, ${B_1}=2$ MHz, and ${\Gamma_1} = 0.5$ MHz.}
\label{pcfandomegaz}
\end{figure}

With the output single-photon pulse given by Eq.~(\ref{aoujtttttt}), we assume ${B_1}=2$ MHz and damped rate ${\Gamma_1}=0.5$ MHz shown in Fig.~\ref{ALPHAOUT2}(a)(b), which denote ${\left| {{\alpha _{ou{t_1}}}(t)} \right|^2}$ and $| {\Omega (t)/({g_{c}}\sqrt N )} |$ as functions of time $t$. It is found from Fig.~\ref{ALPHAOUT2} that the driving field has different shapes when we control to generate an output   single-photon wavepacket in the Markovian and non-Markovian regimes.
Next we show that with a driving field, one can manipulate and change the characteristics of the photon generation.

Fig.~\ref{pcfandomega} compares the control schemes obtained with and without the
Markovian approximations. To be specific, the population
of the state $\left| c \right\rangle $ in the Markovian approximation is compared with the case without the Markovian approximation as shown in Fig.~\ref{pcfandomega}(a), while Fig.~\ref{pcfandomega}(c) corresponds to the comparison of the real driving field with
and without the Markovian approximations. As shown in Fig.~\ref{pcfandomega}(a)(c), the system exhibits the non-Markovian
dynamics, and we easily see that $\Omega (t)$, ${\Omega _f}(t)$ and ${\rho _c}(t)$, ${\rho _{cf}}(t)$ (the subscript $f$ corresponds to the quantities under the Markovian approximation given by Eq.~(\ref{lastmarkov}))
have apparent differences called the backflowing phenomena when the spectral width of the environment is small ($\lambda _1= 2.31$ MHz), which can be understood by memory effects in the photon emission of the non-Markovian environment. Besides, from Fig.~\ref{pcfandomega}(b)(d), we  see the optimal driving field and population of the state $\left| c \right\rangle $ in the non-Markovian case for a large spectral width $\lambda_1$ are in good agreement with those in the Markovian approximation.
It also confirms that the single-photon wavepackets derived with the Markovian approximation have almost the same results as
those in the non-Markovian regime when the spectral width ${\lambda _1}$ equals $30$ MHz in Fig.~\ref{pcfandomega}(b)(d).

In order to see the continuous influence of the increase of the spectral width on the single-photon generation, we plot $\rho _{c}(t)$ and $\rho _{cf}(t)$
as functions of $t$ and $\lambda _1$ in Fig.~\ref{PcPcf}.
Fig.~\ref{PcPcf}(a)(b) show that the population of the excited
state $\rho _{c}(t)$ and $\rho _{cf}(t)$ have obvious differences
when ${\lambda _1}$ increases from 2 MHz to 10 MHz.
As shown in Fig.~\ref{PcPcf}(c)(d), the Markovian approximation produces almost the same results as the exact case when ${\lambda _1}$ increases from 10 MHz to 20 MHz.
Therefore, Fig.~\ref{PcPcf} has given the validity range of the Markovian approximation.

Before concluding the section, we present a discussion on
complex single-photon wavepacket generations with non-Markovian effects, where the  two non-zero detunings ${\delta_1} = 1$ MHz and ${\delta_2} = 2$ MHz.~We assume that the complex single-photon wavepacket takes ${\alpha _{ou{t_1}}}(t) = {A_1}{e^{ - {\Gamma _1}t}}{\sin ^3}({B_1}t){e^{ - i{c_e}t}}$ with ${A_1} = 2\sqrt {2(36B_1^6{\Gamma _1} + 49B_1^4\Gamma _1^3 + 14B_1^2\Gamma _1^5 + \Gamma _1^7)} /(3\sqrt 5 B_1^3)$ in Fig.~\ref{pcfandomegaz}, which leads to  the modulus and argument for the optimal driving field $\Omega (t)$ to generate a single-photon wavepacket of arbitrary temporal shape from an optical cavity coupled with $N$ three-level atoms in the non-Markovian regime, which can be obtained by Eq.~(\ref{omegadelta}). From Fig.~\ref{pcfandomegaz}, we find that gradual increase of phase $c_e$ for the complex single-photon wavepacket with all the other parameters fixed has huge impacts on the modulus and argument for the optimal driving field.

\section{MULTIPLE non-Markovian ENVIRONMENTS INTERACTing WITH THE ONE-SIDED CAVITY}

In many realistic scenarios,
the quantum system can simultaneously couple with the multiple environments \cite{Zhang8108,Combes20172}.
In this section, we study that the cavity simultaneously interacts with
the multiple non-Markovian input-output fields corresponding to Eq.~(\ref{Htotal}).
When the first single-photon wavepacket generated by the Markovian system is the same as the non-Markovian case, setting all the other spectral widths not equalling the first one will lead to the Markovian system not generating the same multiple single-photon wavepackets as the non-Markovian one, while the spectral widths of the different environments taking the same values can generate this. Now we go into the details.
To generate the multiple single-photon wavepackets from the cavity,
assuming ${\alpha _{i{n_j}}}(t)=0$ with Eq.~(\ref{finallyEq4}), we have ${\alpha _{ou{t_j}}}(t) = \int_{{0}}^t {d\tau } {k_j}(t - \tau ){\beta _b}(\tau )$ and 
\begin{equation}
\begin{aligned}
{\beta _b}(t) =& \frac{{{{\dot \alpha }_{ou{t_1}}}(t) + {\lambda _1}{\alpha _{ou{t_1}}}(t)}}{{{\lambda _1}\sqrt {{\gamma _1}} }}\\
 &\cdots \\
 =& \frac{{{{\dot \alpha }_{ou{t_j}}}(t) + {\lambda _j}{\alpha _{ou{t_j}}}(t)}}{{{\lambda _j}\sqrt {{\gamma _j}} }}&\\
 &\cdots \\
=& \frac{{{{\dot \alpha }_{ou{t_m}}}(t) + {\lambda _m}{\alpha _{ou{t_m}}}(t)}}{{{\lambda _m}\sqrt {{\gamma _m}} }}\\
 &\cdots \\
=& \frac{{{{\dot \alpha }_{ou{t_M}}}(t) + {\lambda _M}{\alpha _{ou{t_M}}}(t)}}{{{\lambda _M}\sqrt {{\gamma _M}} }},
\label{beltabM}
\end{aligned}
\end{equation}
for the non-Markovian case, where $j=1,2,{\cdot\cdot\cdot}, M$ and $m=1,2,{\cdot\cdot\cdot}, M$. Eq.~(\ref{beltabM}) shows that the generated single-photon wavepackets are not independent of each other but satisfy certain correlations related to non-Markovian spectral parameters.

In particular, the Markovian approximation gives
\begin{equation}
\begin{aligned}
{\beta _{bf}}(t) &= \frac{{{\alpha _{ou{t_{1f}}}}(t)}}{{\sqrt {{\gamma _1}} }} =  \cdots  =
\frac{{{\alpha _{ou{t_{jf}}}}(t)}}{{\sqrt {{\gamma _j}} }} \\
&=  \cdots  =
\frac{{{\alpha _{ou{t_{mf}}}}(t)}}{{\sqrt {{\gamma _m}} }}=\cdots  =
\frac{{{\alpha _{ou{t_{Mf}}}}(t)}}{{\sqrt {{\gamma _M}} }},
\label{beltabfM}
\end{aligned}
\end{equation}
where ${\beta _{bf}}(t)$ and $\alpha _{ou{t_{jf}}}(t)$ with the subscript $f$ correspond to the quantities under the Markovian approximation.

We discuss the multiple single-photon generations in the Markovian and non-Markovian systems from the following aspects.

(1) If the cavity only interacts with an input-output field ($M=1$), the Markovian system can  generate  the same single-photon wavepacket as the non-Markovian one, i.e., $\alpha _{ou{t_{1f}}}(t)=\alpha _{ou{t_{1}}}(t)$ (the initial conditions ${\alpha _{ou{t_1}}}(0) = 0$, ${\dot \alpha _{ou{t_1}}}(0) = 0$, and ${\ddot \alpha _{ou{t_1}}}( 0 ) = 0$ need to be satisfied, which can be seen from the first paragraph of Sec.~IV for the more details),  where the optimal driving field $\Omega (t)$ of  Eq.~(\ref{OMEGA})  falls in the non-Markovian regime, while the optimal driving field ${\Omega _f}(t)$ under the  Markovian approximation is given by Eq.~(\ref{lastmarkov}).

(2)
Under some special conditions, the Markovian system can generate the same (or different) multiple ($M \ge 2$) single-photon wavepackets as the non-Markovian one.   The seven possible situations should be considered, labeled by (i)-(vii) below.

We assume that the Markovian system can generate a same single-photon wavepacket as the non-Markovian one, e.g.,
\begin{equation}
\begin{aligned}
{\alpha _{ou{t_1}}}(t) &= {\alpha _{ou{t_{1f}}}}(t)\\
 &\equiv \sqrt {\frac{{{\gamma _1}}}{{{\gamma _j}}}}{\alpha _{ou{t_{jf}}}}(t),
\label{abcz21zz}
\end{aligned}
\end{equation}
where the second identity originates from Eq.~(\ref{beltabfM}). Of course, we can also set ${\alpha _{ou{t_1}}}(t) = {\alpha _{ou{t_{pf}}}}(t)$ ($p=2,3,{\cdot\cdot\cdot}, M$), which is discussed in (v), while (vii) corresponds to the case of generating two same single-photon wavepackets ${\alpha _{ou{t_{1f}}}}(t) = {\alpha _{ou{t_{1}}}}(t)$ and ${\alpha _{ou{t_{2f}}}}(t) = {\alpha _{ou{t_{2}}}}(t)$ for $M \ge 3$.
On the contrary, if the Markovian system cannot generate any same single-photon wavepackets as the non-Markovian one, it is not necessary to establish the relationship between Eq.~(\ref{beltabM}) and Eq.~(\ref{beltabfM}).

Substituting
Eq.~(\ref{abcz21zz}) into the first equation of Eq.~(\ref{beltabM}), we get ${\beta _b}(t) = {\dot \alpha _{ou{t_{jf}}}}(t) + {\lambda _1}{\alpha _{ou{t_{jf}}}}(t)/({\lambda _1}\sqrt {{\gamma _j}} )$, which equals the third equation of Eq.~(\ref{beltabM}), then
\begin{equation}
\begin{aligned}
\frac{{{{\dot \alpha }_{ou{t_{jf}}}}(t) + {\lambda _1}{\alpha _{ou{t_{jf}}}}(t)}}{{{\lambda _1}\sqrt {{\gamma _j}} }} = \frac{{{{\dot \alpha }_{ou{t_m}}}(t) + {\lambda _m}{\alpha _{ou{t_m}}}(t)}}{{{\lambda _m}\sqrt {{\gamma _m}} }}.
\label{abcz13a}
\end{aligned}
\end{equation}
If $m=j$, Eq.~(\ref{abcz13a}) is reduced to
\begin{equation}
\begin{aligned}
\frac{{{{\dot \alpha }_{ou{t_{jf}}}}(t)}}{{{\lambda _1}}} -
\frac{{{{\dot \alpha }_{ou{t_j}}}(t)}}{{{\lambda _j}}} +
{\alpha _{ou{t_{jf}}}}(t) - {\alpha _{ou{t_j}}}(t) = 0.
\label{abcz2y}
\end{aligned}
\end{equation}

(i) In the first case, we assume
\begin{equation}
\begin{aligned}
{\alpha _{ou{t_{1f}}}}(t){\rm{ = }}&{\alpha _{ou{t_1}}}(t), \cdots ,{\alpha _{ou{t_{jf}}}}(t)
{\rm{ = }}{\alpha _{ou{t_j}}}(t),\\
 &\cdots ,{\alpha _{ou{t_{Mf}}}}(t){\rm{ = }}{\alpha _{ou{t_M}}}(t),
\label{abcz2aa}
\end{aligned}
\end{equation}
and get from Eq.~(\ref{abcz2y})
\begin{equation}
\begin{aligned}
&{\lambda _1}  =  \cdots  = {\lambda _j} =
\cdots  = {\lambda _M} \equiv {\lambda} ,\label{abcz2aba}
\end{aligned}
\end{equation}
where the differences between the Markovian and non-Markovian systems are that the optimal driving fields have different forms, i.e., $\Omega (t)$ is determined by Eq.~(\ref{OMEGA}), while ${\Omega _f}(t)$ takes Eq.~(\ref{lastmarkov}). According to Eq.~(\ref{beltabfM}) and normalization condition $\sum_{j = 1}^M {{\mu _j}}  = 1$ with $ {\mu _j}=\int {dt{{| {{\alpha _{ou{t_{jf}}}}(t)} |}^2}}  $, we obtain the relation between $\gamma_j$ and $\mu_j$ as follows
\begin{eqnarray}
{\gamma _j} = \frac{{{\mu _j}{\gamma _1}}}{{{\mu _1}}}.
\label{summary}
\end{eqnarray}
For the identical output single-photon wavepackets, i.e., ${\alpha _{ou{t_{1f}}}}(t)  =  \cdots  = {\alpha _{ou{t_{jf}}}}(t) = \cdots  = {\alpha _{ou{t_{Mf}}}}(t)$, we get  ${\mu _1} =  \cdots {\mu _j} =  \cdots  = {\mu _M} = 1/M$.

(ii) However, if all the other spectral widths do not equal the first one, i.e.,
\begin{equation}
\begin{aligned}
{\lambda _1} \ne {\lambda _2} ,\cdots, {\lambda _1} \ne {\lambda _j}   ,\cdots, {\lambda _1} \ne {\lambda _M},
\label{abcz2bb}
\end{aligned}
\end{equation}
we have
\begin{equation}
\begin{aligned}
{\alpha _{ou{t_{2f}}}}(t) \ne& {\alpha _{ou{t_2}}}(t), \cdots ,{\alpha _{ou{t_{jf}}}}(t)
 \ne {\alpha _{ou{t_j}}}(t), \\
 &\cdots ,{\alpha _{ou{t_{Mf}}}}(t) \ne {\alpha _{ou{t_M}}}(t),
\label{abcz2bbzz}
\end{aligned}
\end{equation}
which show that the Markovian system cannot generate the same multiple single-photon wavepackets as the non-Markovian one (A single-photon wavepacket generated here is assumed to be equal, e.g.,
${\alpha _{ou{t_{1f}}}}(t)={\alpha _{ou{t_{1}}}}(t)$ in Eq.~(\ref{abcz21zz})). In this case, the relations given by Eq.~(\ref{beltabM})
between ${\alpha _{ou{t_{j}}}}(t)$ and ${\alpha _{ou{t_{1}}}}(t)$ in  the non-Markovian regime are determined as follows
\begin{equation}
\begin{aligned}
&{{\alpha }_{ou{t_j}}}(t) =  \frac{{{\lambda _j}\sqrt {{\gamma _j}} }}{{{\lambda _1}\sqrt {{\gamma _1}} }}\int_0^t {d{t_1}[{{{\dot {\alpha} }_{ou{t_1}}}({t_1}) + {\lambda _1}{\alpha _{ou{t_1}}}({t_1})}]}\\
& \quad \quad \quad \quad \,\,\,\, \times {e^{ - {\lambda _j}\left( {t - {t_1}} \right)}},
\label{alphaoutM}
\end{aligned}
\end{equation}
which satisfies the normalization condition $\sum_{j = 1}^M {{\nu_j}}  = 1$ with
\begin{eqnarray}
 {\nu _j} =\int {dt{{| {{\alpha _{ou{t_{j}}}}(t)} |}^2}}  .\label{summary1}
\label{summary2}
\end{eqnarray}
\begin{figure}
\centering
\includegraphics[height=8.0cm,width=9.1cm]{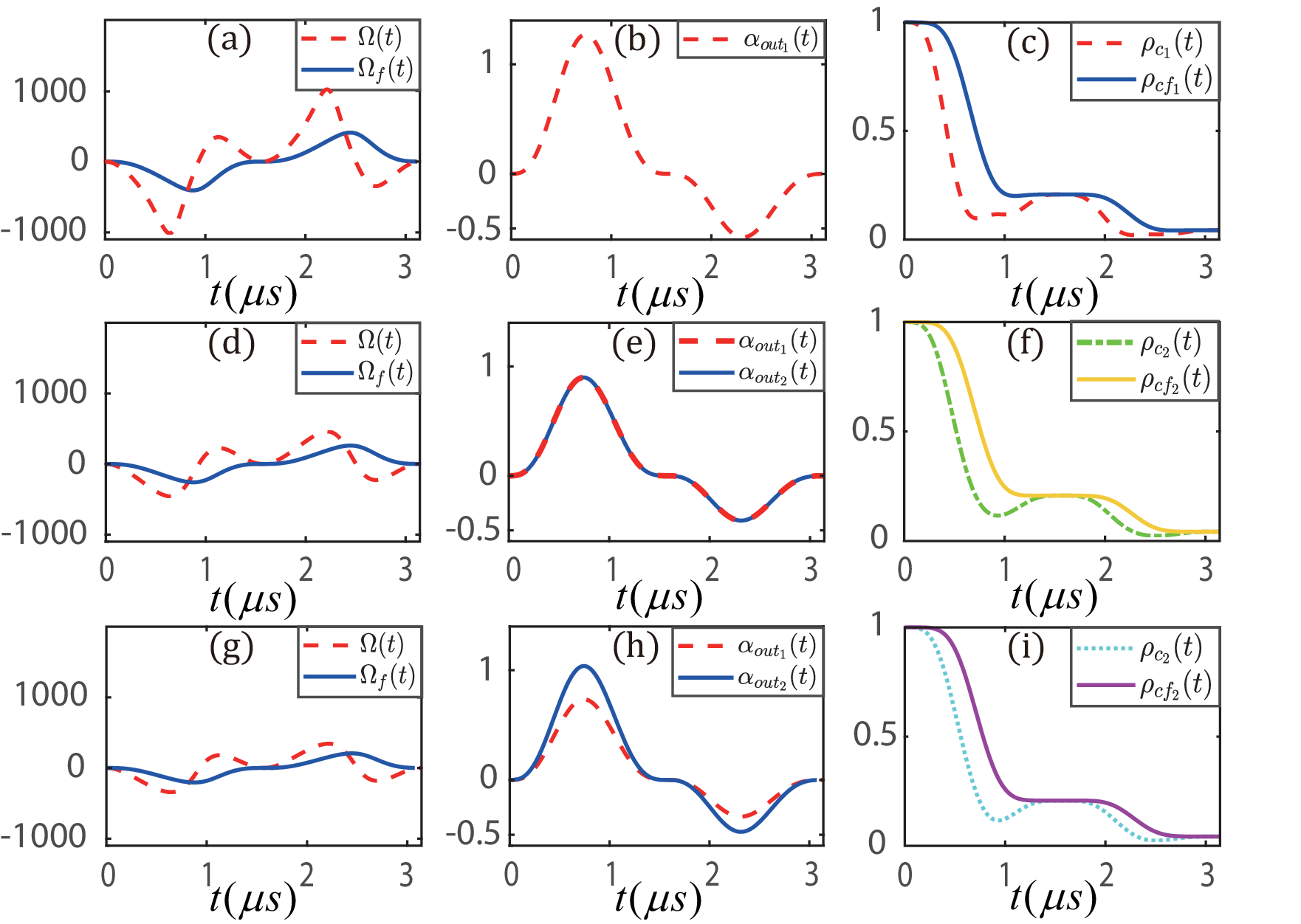}
\caption{(Color online) In order to compare the cavity interacting with one (${\alpha _{ou{t_1}}}(t) = {A_2}{e^{ - {\Gamma _2}t}}{\sin ^3}{B_2}t$,
where ${A_2} = 2\sqrt {2(36B_2^6{\Gamma _2} + 49B_2^4\Gamma _2^3 + 14B_2^2\Gamma _2^5 + \Gamma _2^7)} /(3\sqrt 5 B_2^3)$),
two identical (${\alpha _{ou{t_1}}}(t) = {\alpha _{ou{t_2}}}(t)= A_3{e^{ - \Gamma_3 t}}{\sin ^3} B_3 t$
with ${A_3} = 2\sqrt {36B_3^6{\Gamma _3} + 49B_3^4\Gamma _3^3 + 14B_3^2\Gamma _3^5 + \Gamma _3^7} /(3\sqrt 5 B_3^3)$),
and two different input-output fields (${\alpha _{ou{t_1}}}(t) = {A_4}{e^{ - {\Gamma _4}t}}{\sin ^3}{B_4}t$ and
${{\alpha }_{ou{t_2}}}(t) = \sqrt {{\gamma _2}/{\gamma _1}}  \int_0^t {d{t_1}
[\dot{{\alpha}}_{out_1}({t_1}) + \lambda {{\alpha }_{ou{t_1}}}({t_1})]}
 {e^{ - \lambda \left( {t - {t_1}} \right)}}$ with ${A_4} = 2\sqrt {2(36B_4^6{\Gamma _4}
 + 49B_4^4\Gamma _4^3 + 14B_4^2\Gamma _4^5 + \Gamma _4^7)} /(3\sqrt {15} B_4^3)$)
in the Markovian and non-Markovian cases, we plot the  optimal driving field
${\Omega _f}(t)$ in Eq.~(\ref{lastmarkov}) and $\Omega (t)$ given by Eq.~(\ref{OMEGA}) with and without the Markovian approximations
in (a)(d)(g), output wavepackets in (b)(e)(h), and population (${\rho _c}(t)$ in  Eq.~(\ref{A})
and ${\rho _{cf}}(t)$ in Eq.~(\ref{lastmarkov})) in (c)(f)(i). The parameters chosen are ${\gamma _1} = 10$ MHz, ${\gamma '} = 6\pi$ MHz, ${g_{c}} = 30\pi$ MHz,
${\lambda} = 2.31$ MHz, ${\delta _1} = {\delta _2}=0$, $N = 40$, ${B_2} = {B_3} = {B_4} = 2$ MHz, and ${\Gamma _2} = {\Gamma _3} = {\Gamma _4} = 0.5$ MHz.}
\label{anotheralphaout2}
\end{figure}
\begin{figure}
\centering
\includegraphics[height=8.0cm,width=9.1cm]{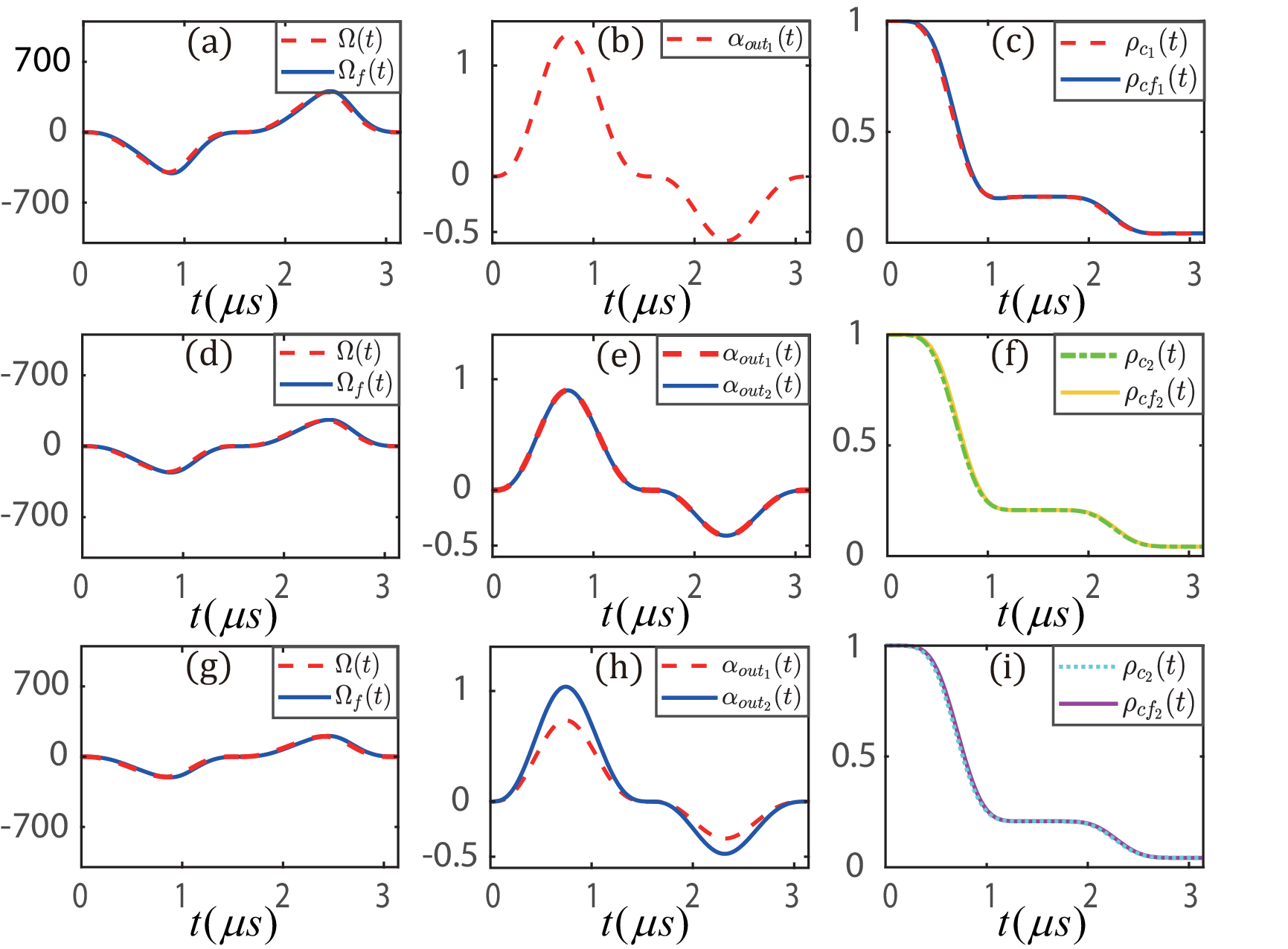}
\caption{(Color online) The figure shows the case of the Markovian approximation, where the parameters chosen are same as Fig.~\ref{anotheralphaout2} except $\lambda = 30$ MHz.}
\label{anotheralphaoutomega30}
\end{figure}
\begin{figure}
\centering
\includegraphics[height=8.0cm,width=9.1cm]{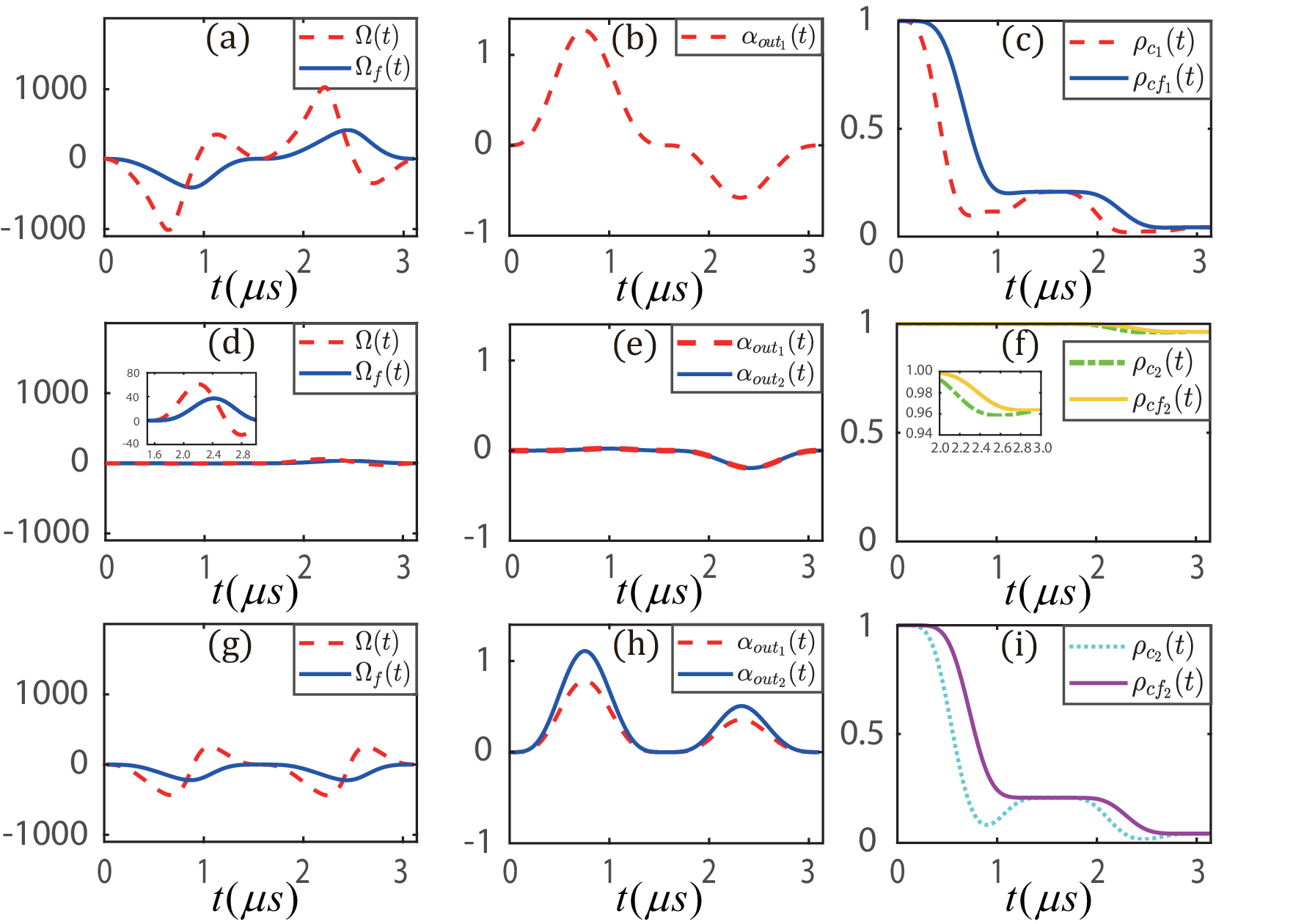}
\caption{(Color online) The three rows correspond to the three different output  single-photon wavepackets with non-Markovian effects, i.e., (b) ${\alpha _{ou{t_1}}}(t) = {A_2}{e^{ - {\Gamma _2}t}}{\sin ^3}{B_2}t$ with ${A_2} = 2\sqrt {2(36B_2^6{\Gamma _2} + 49B_2^4\Gamma _2^3 + 14B_2^2\Gamma _2^5 + \Gamma _2^7)} /(3\sqrt 5 B_2^3)$, (e) ${\alpha _{ou{t_1}}}(t) = {\alpha _{ou{t_2}}}(t) = {A_3}{t^3}{e^{ - {\Gamma _3}t}}{\sin ^3}{B_3}t$ with ${A_3} = 2\sqrt 2 /[45\sqrt {1/(72\Gamma _3^7) + {\Gamma _3}({d_1} + {d_2} + 240B_3^4{d_3} + 72B_3^2{d_4})/720} ]$, where ${d_1} = 6/{(4B_3^2 + \Gamma _3^2)^4} - 1/{(9B_3^2 + \Gamma _3^2)^4} - 15/{(B_3^2 + \Gamma _3^2)^4}$, ${d_2} = 192B_3^6[5/{(B_3^2 + \Gamma _3^2)^7} - 128/{(4B_3^2 + \Gamma _3^2)^7} + 243/{(9B_3^2 + \Gamma _3^2)^7}]$, ${d_3} = 32/{(4B_3^2 + \Gamma _3^2)^6} - 27/{(9B_3^2 + \Gamma _3^2)^6} - 5/{(B_3^2 + \Gamma _3^2)^6}$, ${d_4} = 5/{(B_3^2 + \Gamma _3^2)^5} - 8/{(4B_3^2 + \Gamma _3^2)^5} + 3/{(9B_3^2 + \Gamma _3^2)^5}$, and (h) ${ \alpha _{ou{t_1}}}(t) = {A_4}{e^{ - {\Gamma _4}t}}{\sin ^4}{B_4}t$~with~${A_4} = 2\sqrt {576B_4^8{\Gamma _4} + 820B_4^6\Gamma _4^3 + 273B_4^4\Gamma _4^5 + 30B_4^2\Gamma _4^7 + \Gamma _4^9} /z$, $z = 3\sqrt {105} B_4^4$, ${\alpha _{ou{t_2}}}(t) = \sqrt {{\gamma _2}/{\gamma _1}} \int_0^t {d{t_1}[{{\mathop \alpha \limits^. }_{ou{t_1}}}({t_1}) + \lambda {\alpha _{ou{t_1}}}({t_1})]}  \times {e^{ - \lambda \left( {t - {t_1}} \right)}}$. In this case, we take ${B_2} = {B_3} = {B_4} = 2$ MHz, ${\Gamma_2}={\Gamma_3}={\Gamma_4}=0.5$ MHz, and $\lambda = 2.31$ MHz. Other parameters and vertical ordinates are given in Fig.~\ref{anotheralphaout2}.}
\label{fangfa211}
\end{figure}
\begin{figure}
\centering
\includegraphics[height=8.0cm,width=9.1cm]{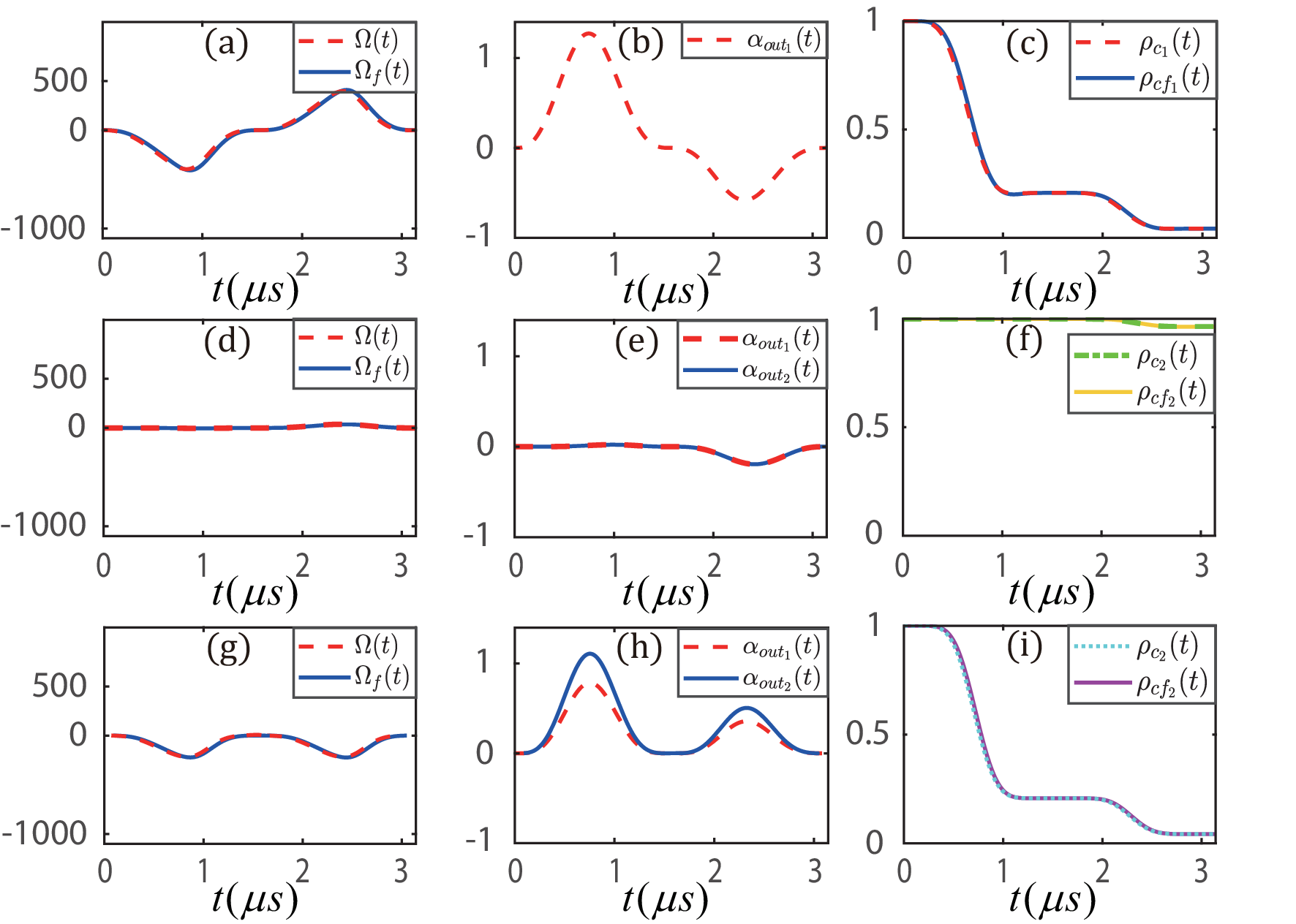}
\caption{(Color online) The situation discussed in the figure corresponds to the Markovian approximation with ${\lambda}=30$ MHz, where the other parameters are the same as Fig.~\ref{fangfa211}.}
\label{fangfa212}
\end{figure}
\begin{figure}
\centering
\includegraphics[height=6.2cm,width=8.5cm]{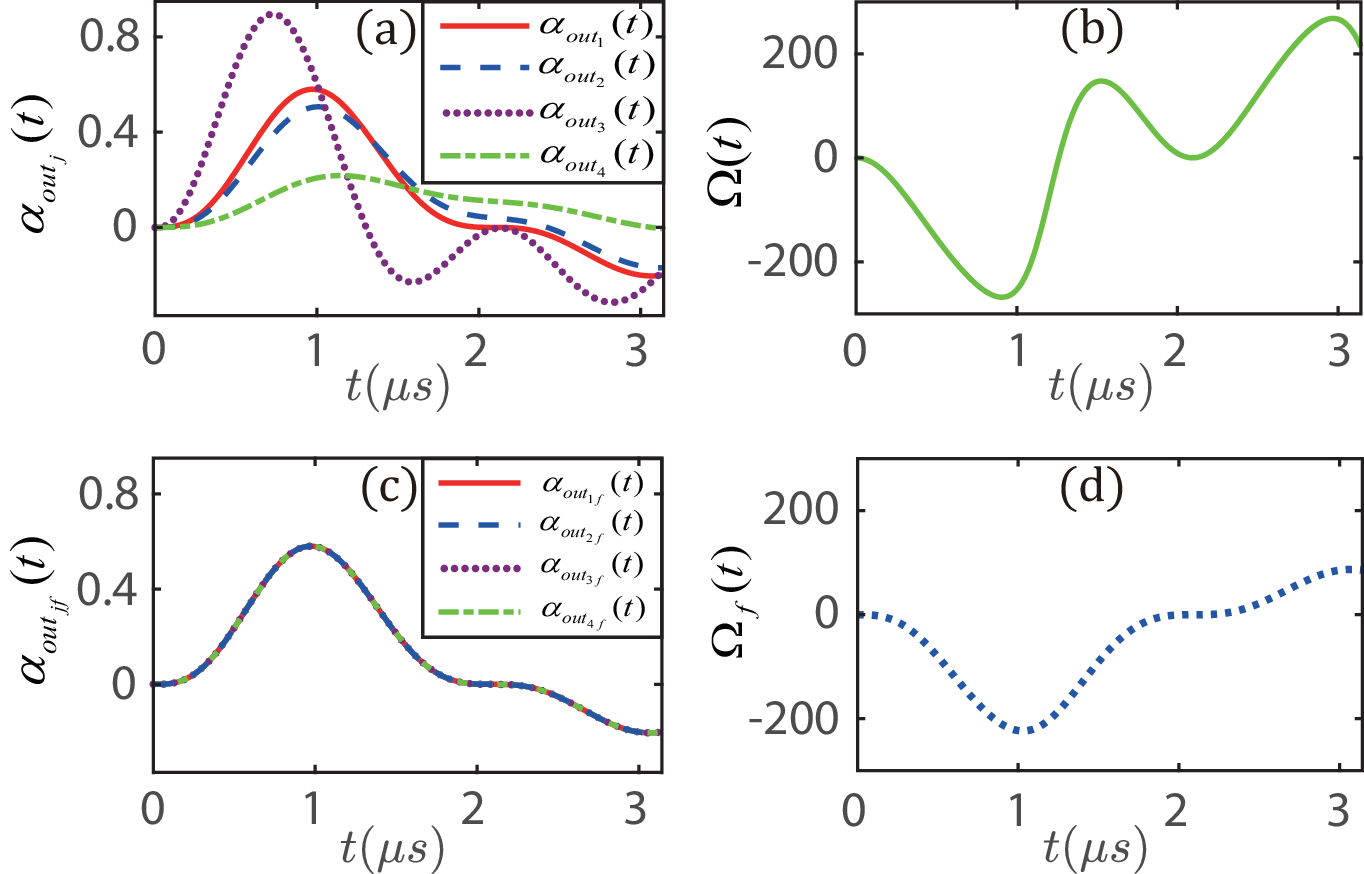}
\caption{(Color online) To generate the multiple ($M=4$) single-photon wavepackets with the normalization factors $\nu_1=1/4$, $\nu_2=1/5$, $\nu_3=1/2$, and $\nu_4=1/20$, we plot four output single-photon wavepackets ${\alpha _{ou{t_1}}}(t)$, ${\alpha _{ou{t_2}}}(t)$, ${\alpha _{ou{t_3}}}(t)$, and ${\alpha _{ou{t_4}}}(t)$ given by Eq.~(\ref{bezsez}) in the non-Markovian regime in Fig.~\ref{NM_output_omega}(a), which are separated, while four output single-photon wavepackets ${\alpha _{ou{t_{1f}}}}(t)$, ${\alpha _{ou{t_{2f}}}}(t)$, ${\alpha _{ou{t_{3f}}}}(t)$, and ${\alpha _{ou{t_{4f}}}}(t)$  under the Markovian approximation in Fig.~\ref{NM_output_omega}(c)  are totally overlapped due to the equal normalization factors $\mu_1=\mu_2=\mu_3=\mu_4=1/4$ revealed from Eq.~(\ref{bezsez11c}). $\Omega (t)$ and $\Omega_f (t)$ are determined by Eqs.~(\ref{OMEGA}) and~(\ref{lastmarkov}) in Fig.~\ref{NM_output_omega}(b) and (d), respectively. The other parameters chosen are ${\lambda_1} = 2$ MHz, $B=1.5$ MHz, $\Gamma=0.5$ MHz, ${\gamma _1} = 10$ MHz, ${\gamma '} = 6\pi$ MHz, ${g_{c}} = 30\pi$ MHz, ${\delta _1} = {\delta _2}=0$, and $N = 40$. Through these parameters, we obtain ${\lambda_2} = 1.52$ MHz, ${\lambda_3} = 24.87$ MHz, and ${\lambda_4} = 0.439$ MHz given by Eq.~(\ref{bezsez11}).}
\label{NM_output_omega}
\end{figure}

(iii) As can be seen from Eq.~(\ref{abcz2y}), the generated single-photon wavepackets can also be partially equal, e.g., ${\alpha _{ou{t_{1f}}}}(t){\rm{ = }}{\alpha _{ou{t_1}}}(t)$, ${\alpha _{ou{t_{3f}}}}(t){\rm{ = }}{\alpha _{ou{t_3}}}(t)$, ${\alpha _{ou{t_{5f}}}}(t){\rm{ = }}{\alpha _{ou{t_5}}}(t)=\cdots$ with ${\lambda _1}  =   {\lambda _3}=   {\lambda _5}=\cdots$, and ${\alpha _{ou{t_{2f}}}}(t){\rm{ \ne }}{\alpha _{ou{t_2}}}(t)$, ${\alpha _{ou{t_{4f}}}}(t){\rm{ \ne }}{\alpha _{ou{t_4}}}(t)$, ${\alpha _{ou{t_{6f}}}}(t){\rm{ \ne }}{\alpha _{ou{t_6}}}(t)=\cdots$ with ${\lambda _1} \ne {\lambda _2},{\lambda _1} \ne {\lambda _4} ,{\lambda _1} \ne {\lambda _6}\cdots$.

(iv)~Assuming $m\ne j$ and   ${\alpha _{ou{t_{jf}}}}(t){\rm{ = }}{\alpha _{ou{t_m}}}(t)$, based on ${\alpha _{ou{t_1}}}(t) = {\alpha _{ou{t_{1f}}}}(t)$, Eq.~(\ref{abcz13a}) leads to $\lambda_1=\lambda_m$ and $\gamma_j=\gamma_m$. For  $\lambda_1\ne\lambda_m$ or $\gamma_j\ne\gamma_m$, we have ${\alpha _{ou{t_{jf}}}}(t)\ne{\alpha _{ou{t_m}}}(t)$.

(v) If ${\alpha _{ou{t_1}}}(t){\rm{ }} = {\alpha _{ou{t_{pf}}}}(t)$ ($p=2,3,{\cdot\cdot\cdot}, M$), substituting it into Eq.~(\ref{beltabM}), we get $[{{\dot \alpha }_{ou{t_{pf}}}}(t) + {\lambda _1}{\alpha _{ou{t_{pf}}}}(t)]/({\lambda _1}\sqrt {{\gamma _1}} ) = [{{\dot \alpha }_{ou{t_m}}}(t) + {\lambda _m}{\alpha _{ou{t_m}}}(t)]/({\lambda _m}\sqrt {{\gamma _m}} )$, which gives  $\lambda_1=\lambda_m$ and $\gamma_1=\gamma_m$ when ${\alpha _{ou{t_{pf}}}}(t){\rm{ = }}{\alpha _{ou{t_m}}}(t)$. Setting $\lambda_1\ne\lambda_m$ or $\gamma_1\ne\gamma_m$, we obtain ${\alpha _{ou{t_{pf}}}}(t)\ne{\alpha _{ou{t_m}}}(t)$ when ${\alpha _{ou{t_1}}}(t) = {\alpha _{ou{t_{pf}}}}(t)$.

(vi) When the same single-photon wavepacket generated by the Markovian and non-Markovian systems in Eq.~(\ref{abcz21zz}) is not assumed to be ${\alpha _{ou{t_1}}}(t) = {\alpha _{ou{t_{1f}}}}(t)$ but ${\alpha _{ou{t_j}}}(t) = {\alpha _{ou{t_{jf}}}}(t)$, we have $\lambda_j=\lambda_m$ if ${\alpha _{ou{t_{mf}}}}(t){\rm{ = }}{\alpha _{ou{t_m}}}(t)$, then ${\alpha _{ou{t_{mf}}}}(t){\rm{ \ne }}{\alpha _{ou{t_m}}}(t)$ for $\lambda_j\ne\lambda_m$.

(vii) If the Markovian system
can generate two same single-photon wavepackets ${\alpha _{ou{t_{1f}}}}(t) = {\alpha _{ou{t_{1}}}}(t)$ and ${\alpha _{ou{t_{2f}}}}(t) = {\alpha _{ou{t_{2}}}}(t)$ (leading to $\lambda_1=\lambda_2$) as the non-Markovian one (corresponding to the case for generating a same single-photon wavepacket in Eq.~(\ref{abcz21zz})) for $M \ge 3$, Eqs.~(\ref{abcz2aa}) and (\ref{abcz2aba}) remain unchanged, while Eqs.~(\ref{abcz2bb}) and (\ref{abcz2bbzz}) become ${\lambda _1}{\rm{ = }}{\lambda _2} \ne {\lambda _3}, \cdots ,{\lambda _1}{\rm{ = }}{\lambda _2} \ne {\lambda _j}, \cdots ,{\lambda _1}{\rm{ = }}{\lambda _2} \ne {\lambda _M}$ and ${\alpha _{ou{t_{3f}}}}(t) \ne {\alpha _{ou{t_3}}}(t), \cdots , {\alpha _{ou{t_{jf}}}}(t) \ne {\alpha _{ou{t_j}}}(t), \cdots ,{\alpha _{ou{t_{Mf}}}}(t) \ne {\alpha _{ou{t_M}}}(t)$.
\begin{figure}
\centering
\includegraphics[height=6.2cm,width=8.5cm]{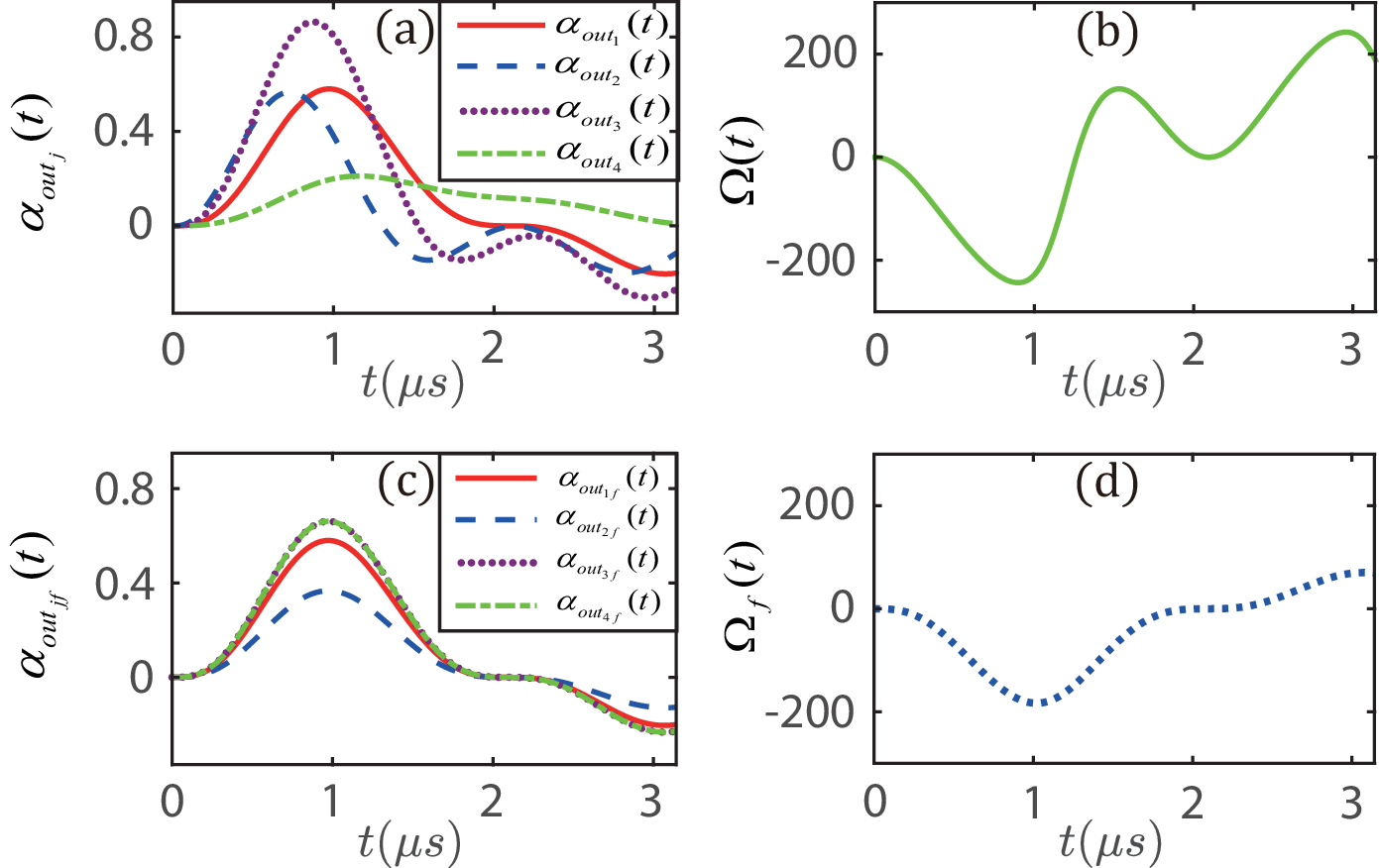}
\caption{(Color online) With the partially non-equal normalization factors $\mu_1=1/4$, $\mu_2=1/10$, $\mu_3=13/40$, and $\mu_4=13/40$ ($M=4$), two output single-photon wavepackets ${\alpha _{ou{t_{3f}}}}(t)$ and ${\alpha _{ou{t_{4f}}}}(t)$ in Eq.~(\ref{bezsez11c}) under the Markovian approximation are overlapped shown in Fig.~\ref{Mar_output_omega}(c), while four output single-photon wavepackets  ${\alpha _{ou{t_1}}}(t)$, ${\alpha _{ou{t_2}}}(t)$, ${\alpha _{ou{t_3}}}(t)$, and ${\alpha _{ou{t_4}}}(t)$ of   Eq.~(\ref{bezsez}) in the non-Markovian regime are separated in Fig.~\ref{Mar_output_omega}(a), where the normalization factors $\nu_1=1/4$, $\nu_2=1/5$, $\nu_3=1/2$, and $\nu_4=1/20$. $\Omega (t)$ and $\Omega_f (t)$ take Eqs.~(\ref{OMEGA}) and~(\ref{lastmarkov}) in Fig.~\ref{Mar_output_omega}(b) and (d), respectively. The other parameters chosen are ${\lambda_1} = 2$ MHz, $B=1.5$ MHz, $\Gamma=0.5$ MHz, ${\gamma _1} = 10$ MHz, ${\gamma '} = 6\pi$ MHz, ${g_{c}} = 30\pi$ MHz, ${\delta _1} = {\delta _2}=0$, and $N = 40$, which
lead to ${\lambda_2} = 24.87$ MHz, ${\lambda_3} = 4.3$ MHz, and ${\lambda_4} = 0.36$ MHz calculated by Eq.~(\ref{bezsez11}).}
\label{Mar_output_omega}
\end{figure}
\begin{figure}
\centering
\includegraphics[height=6.2cm,width=8.5cm]{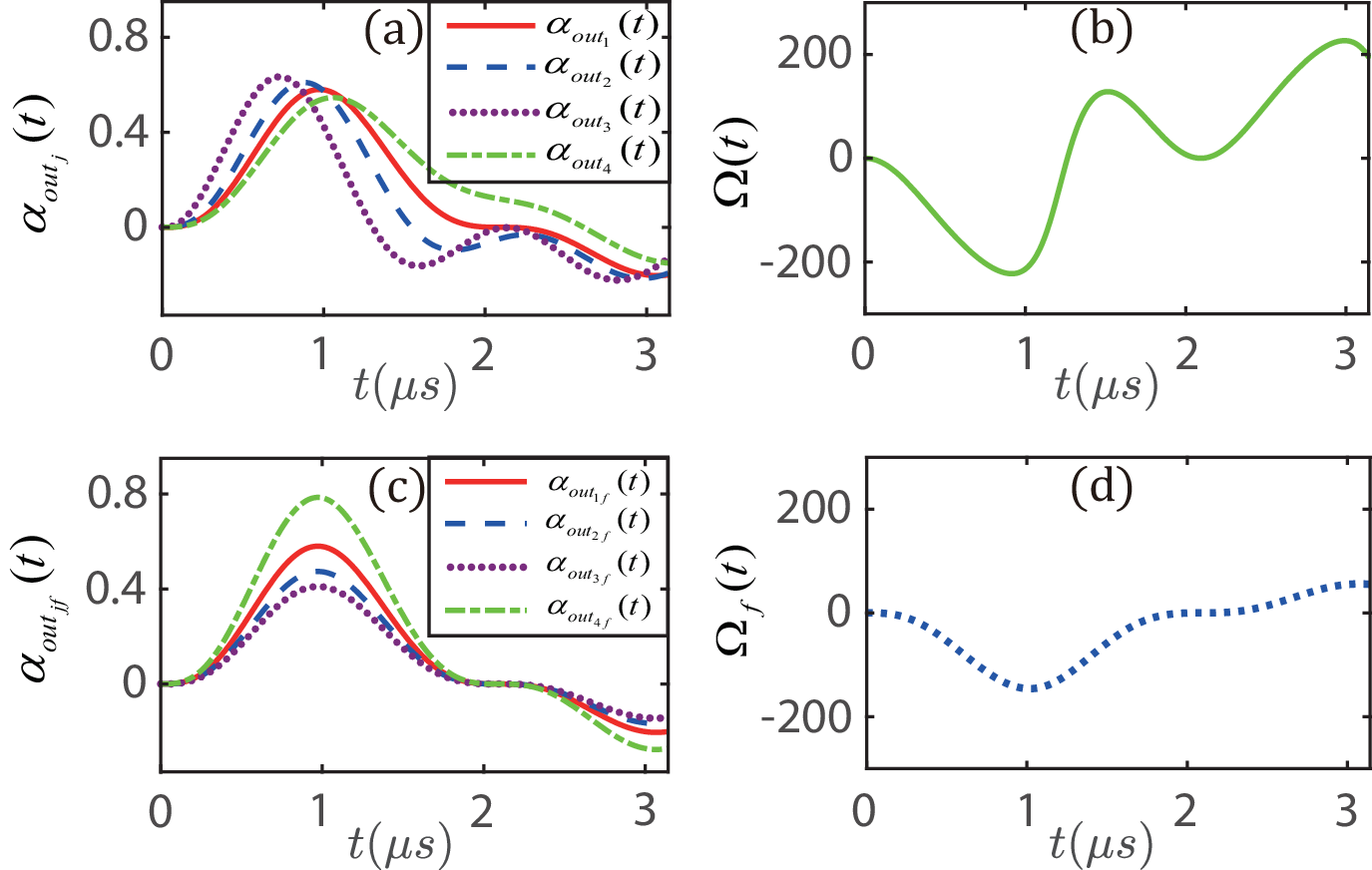}
\caption{(Color online) The completely non-equal normalization factors $\mu_1=1/4$, $\mu_2=1/6$, $\mu_3=1/8$, and $\mu_4=11/24$ ($M=4$) lead to that four output single-photon wavepackets ${\alpha _{ou{t_{1f}}}}(t)$, ${\alpha _{ou{t_{2f}}}}(t)$, ${\alpha _{ou{t_{3f}}}}(t)$, and ${\alpha _{ou{t_{4f}}}}(t)$ given by Eq.~(\ref{bezsez11c}) under the Markovian approximation are completely separated in Fig.~\ref{NM_Mar_output_Omega}(c), while four output single-photon wavepackets in the non-Markovian regime given by ${\alpha _{ou{t_1}}}(t)$, ${\alpha _{ou{t_2}}}(t)$, ${\alpha _{ou{t_3}}}(t)$, and ${\alpha _{ou{t_4}}}(t)$ of Eq.~(\ref{bezsez}) are shown in Fig.~\ref{NM_Mar_output_Omega}(a).
$\Omega (t)$ and $\Omega_f (t)$ are determined by Eqs.~(\ref{OMEGA})  and (\ref{lastmarkov}) in Fig.~\ref{NM_Mar_output_Omega}(b) and (d), respectively.
The other parameters chosen are $\nu_1=\nu_2=\nu_3=\nu_4=1/4$, ${\lambda_1} = 2$ MHz, $B=1.5$ MHz, $\Gamma=0.5$ MHz, ${\gamma _1} = 10$ MHz, ${\gamma '} = 6\pi$ MHz, ${g_{c}} = 30\pi$ MHz, ${\delta _1} = {\delta _2}=0$, and $N = 40$. With these parameters, Eq.~(\ref{bezsez11}) gives ${\lambda_2} = 4.047$ MHz, ${\lambda_3} = 24.87$ MHz, and ${\lambda_4} = 1.028$ MHz.}
\label{NM_Mar_output_Omega}
\end{figure}

With the above similar discussions, the results in (i)-(vii) can also be obtained by Eqs.~(\ref{beltabfM}) and (\ref{alphaoutM}), which especially lead to  ${\alpha _{ou{t_j}}}(t)/{\alpha _{ou{t_1}}}(t) = \sqrt {{\gamma _j}/{\gamma _1}} $ if ${\lambda _1} = {\lambda _j}$ (also derived by ${\alpha _{ou{t_j}}}(t) = {\lambda _j}\sqrt {{\gamma _j}} \int_0^t {d\tau } {e^{ - {\lambda _j}(t - \tau )}}{\beta _b}(\tau )$ above Eq.~(\ref{beltabM})). Next, we are not going to discuss the situations (iii)-(vii) in detail, but mainly focus on the cases (i) and (ii).

\subsection{Multiple single-photon generations by setting the
equal spectral widths for the different environments}
If the spectral widths of the different environments take the same values in Eq.~(\ref{abcz2aba}), we show that the system under the Markovian approximation can generate the same single-photon wavepackets as the non-Markovian one shown in Figs.~\ref{anotheralphaout2}-\ref{fangfa212},  i.e., $\alpha _{ou{t_{jf}}}(t)=\alpha _{ou{t_{j}}}(t)$ given by  Eq.~(\ref{abcz2aa}).
We consider three types of working mechanisms:
(I) the one-sided cavity interacts with one input-output field ($M=1$ and $\mu_1= 1$);
(II) the one-sided cavity interacts simultaneously with two identical input-output fields  ($M=2$, $\lambda_1=\lambda_2 \equiv \lambda$, and $\mu_1=\mu_2= 1/2$);
(III) the one-sided cavity interacts simultaneously with two different input-output fields ($M=2$, $\lambda_1=\lambda_2 \equiv \lambda$, $\mu_1= 1/3$, and $\mu_2= 2/3$).
We plot the system that works in three working mechanisms in Fig.~\ref{anotheralphaout2}, which shows the driving field,
the output fields, the population of
state $| c \rangle$ in the non-Markovian and Markovian cases with
a comparison when the parameter ${\lambda}$ is fixed to the value $2.31$ MHz in three cases.
Fig.~\ref{anotheralphaout2}(a)(d)(g) show the control driving fields obtained
with and without the Markovian approximations are different in three cases.
Fig.~\ref{anotheralphaout2}(b)(e)(h) correspond to the shape of an output single-photon wavepacket, two identical
wavepackets, and two different wavepackets, respectively.
Fig.~\ref{anotheralphaout2}(c)(f)(i) show the difference of the
population of the state $| c \rangle $
when the system works in three cases.
Compared with the Markovian case,
we can learn that non-Markovianity caused backflow to the state $| c \rangle$ occurs when
the spectral width $\lambda$ is small in three cases.

However, when the spectral width is tuned to $\lambda =30$ MHz, the results given by the  non-Markovian regime are in good agreement with those given in the Markovian approximation.
In Fig.~\ref{anotheralphaoutomega30}(a)(d)(g), the Markovian approximation produces almost the same driving field results
as the exact solution (non-Markovian regime with $\lambda =30$ MHz).
Similarly, the population of the state $\left| c \right\rangle $ in the non-Markovian regime
controlled by the driving field is consistent with that in the Markovian
approximation in three cases when ${\lambda} = 30$ MHz in Fig.~\ref{anotheralphaoutomega30}(c)(f)(i).

As the second concrete example, we take different function forms of output field envelops
in Figs.~\ref{fangfa211} and~\ref{fangfa212}, where the three rows correspond respectively
to ${\alpha _{ou{t_1}}}(t) = {A_2}{e^{ - {\Gamma _2}t}}{\sin ^3}{B_2}t$,
${\alpha _{ou{t_1}}}(t) = {\alpha _{ou{t_2}}}(t) = {A_3}{t^3}{e^{ - {\Gamma_3}t}}{\sin ^3}
{B_3}t$, ${ \alpha _{ou{t_1}}}(t) = {A_4}{e^{ - {\Gamma_4}t}}{\sin ^4}{B_4}t$ and
${{ \alpha }_{ou{t_2}}}(t) = \sqrt {{\gamma _2}/{\gamma _1}}  \int_0^t {d{t_1}
[\dot{{\alpha}}_{out_1}({t_1}) + \lambda {{ \alpha }_{ou{t_1}}}({t_1})]}
\times {e^{ - \lambda \left( {t - {t_1}} \right)}}$. The parameters chosen are
${B_2}={B_3}={B_4}=2$ MHz and ${\Gamma_2}={\Gamma_3}={\Gamma_4}$=0.5 MHz.
We show the output field envelopes are obviously different from those of
Fig.~\ref{anotheralphaout2}. Similarly, the difference between Fig.~\ref{fangfa211}
and Fig.~\ref{fangfa212} depends on the value of spectral width $\lambda$,
where ${\lambda}=2.31$ MHz corresponds to the non-Markovian regime, as shown
in Fig.~\ref{fangfa211}, while the Markovian approximation is characterized
by ${\lambda}=30$ MHz in Fig.~\ref{fangfa212}.

\subsection{Multiple single-photon generations for all the other spectral widths not equalling the first one}

\begin{figure}
\centering
\includegraphics[height=6.2cm,width=8.5cm]{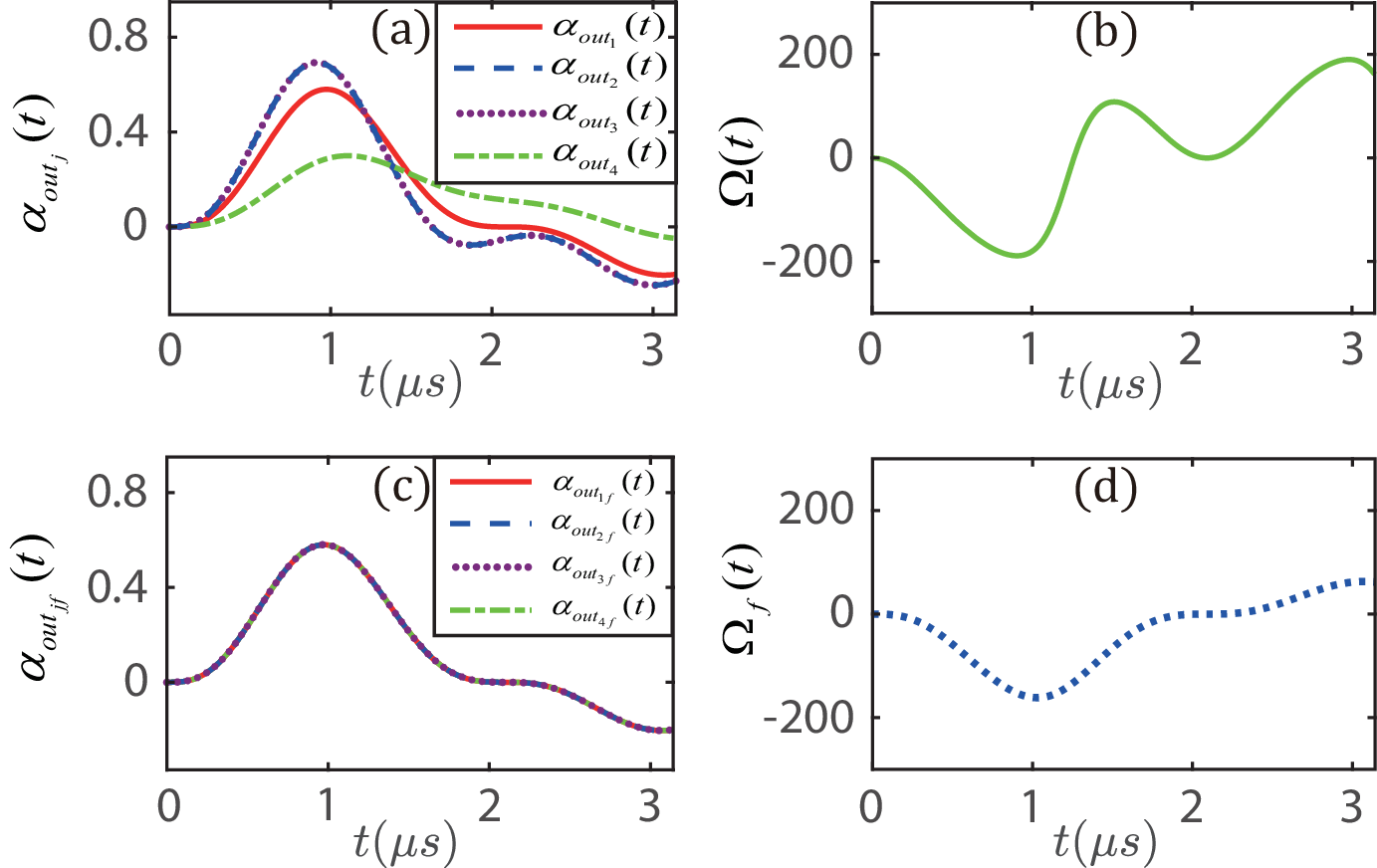}
\caption{(Color online) The figure shows that two equal single-photon wavepackets ${\alpha _{ou{t_2}}}(t)$ and ${\alpha _{ou{t_3}}}(t)$ of four single-photon wavepackets in  the non-Markovian regime are generated in Fig.~\ref{NM_Mar_output_Omegaz1}(a), where  ${\lambda_2} ={\lambda_3} = 3.149$ MHz, $\mu_2=\mu_3=1/4$, and $\nu_2=\nu_3=1/3$ satisfy Eq.~(\ref{czx1}). However, four output single-photon wavepackets ${\alpha _{ou{t_{1f}}}}(t)$, ${\alpha _{ou{t_{2f}}}}(t)$, ${\alpha _{ou{t_{3f}}}}(t)$, and ${\alpha _{ou{t_{4f}}}}(t)$ given by Eq.~(\ref{bezsez11c}) under the Markovian approximation in Fig.~\ref{NM_Mar_output_Omegaz1}(c)  are totally overlapped.
The parameters chosen are $\mu_1=\mu_4=1/4$, $\nu_1=1/4$, $\nu_4=1/12$, and ${\lambda_4} = 0.667$ MHz, which meet Eq.~(\ref{bezsez11}). The other parameters are the same as Fig.~\ref{NM_output_omega}.}
\label{NM_Mar_output_Omegaz1}
\end{figure}

If the spectral widths satisfy Eq.~(\ref{abcz2bb}), we take the first output single-photon wavepacket as ${\alpha _{ou{t_1}}}(t) =E_1{e^{ - {\Gamma}t}}{\sin ^3}{B}t$, and then obtain ${{\alpha }_{ou{t_j}}}(t)$ from Eq.~(\ref{alphaoutM}) as follows
\begin{equation}
\begin{aligned}
{{\alpha }_{ou{t_j}}}(t)  = {D_j}{e^{ -{\lambda _j} t}}\left( {24{B^3}\frac{{{\lambda _1} - {\lambda _j}}}{{{C_j}}} + 3{h_1^{(j)}} - {h_3^{(j)}}} \right),
\label{bezsez}
\end{aligned}
\end{equation}
where ${D_j} = E_1{\lambda _j}\sqrt {{\gamma _j}/{\gamma _1}} /(4{\lambda _1})$, $E_1= {2\sqrt {2\nu_1(36B^6{\Gamma } +
49B^4\Gamma ^3 + 14B^2\Gamma ^5 + \Gamma ^7)} /(3\sqrt 5 B^3)}$, ${h_n^{(j)}} = {e^{t({\lambda _j} -\Gamma)}}\{ nB({\lambda _j} - {\lambda _1})\cos (nBt) +
[{(nB)^2} + (\Gamma- {\lambda _1})(\Gamma- {\lambda _j})]\sin (nBt)\} /[{(nB)^2} +
{(\Gamma- {\lambda _j})^2}]$, and ${C_j} = [{B^2} + {(\Gamma- {\lambda _j})^2}][9{B^2} +
{(\Gamma- {\lambda _j})^2}]$. Similar to Eq.~(\ref{summary}), Eq.~(\ref{summary1})  gives
\begin{equation}
\begin{aligned}
{\gamma _j} = \frac{{5{\nu _j}{\gamma _1}\lambda _1^2[{B^2} + {{(\Gamma  + {\lambda _j})}^2}][9{B^2} + {{(\Gamma  + {\lambda _j})}^2}]}}{{{\lambda _j}{\nu _1}[16\Gamma (4{B^2} + {\Gamma ^2})\lambda _1^2 + {a_1}{\lambda _j} + 4\Gamma {a_2}\lambda _j^2 + {a_2}\lambda _j^3]}},
\label{bezsez11x}
\end{aligned}
\end{equation}
where ${a_1} = 5({B^2} + {\Gamma^2})(9{B^2} + {\Gamma^2}) + (41{B^2} + 29{\Gamma^2})\lambda _1^2$ and
${a_2} = 9{B^2} + {\Gamma^2} + 5\lambda _1^2$. We show that the decay rate $\gamma_j$ in Eq.~(\ref{beltabM}) equals $\gamma_j$ given by Eq.~(\ref{beltabfM}), together with Eqs.~(\ref{summary}) and (\ref{bezsez11x}), which lead to $M$ constraint conditions
\begin{equation}
\begin{aligned}
\frac{{5{\nu _j}{\mu _1}\lambda _1^2}}{{{\mu _j}{\lambda _j}{\nu _1}}} = \frac{{16\Gamma (4{B^2} + {\Gamma ^2})\lambda _1^2 + {a_1}{\lambda _j} + 4\Gamma {a_2}\lambda _j^2 + {a_2}\lambda _j^3}}{{[{B^2} + {{(\Gamma  + {\lambda _j})}^2}][9{B^2} + {{(\Gamma  + {\lambda _j})}^2}]}}.
\label{bezsez11}
\end{aligned}
\end{equation}
We show that the equality of different spectral widths is not restricted (e.g., see ${\lambda_2} ={\lambda_3} = 3.149$ MHz $\ne {\lambda _1}$ in Fig.~\ref{NM_Mar_output_Omegaz1}(a)) as long as they satisfy Eqs.~(\ref{abcz2bb}) and (\ref{bezsez11}) when all the other parameters are fixed (If $j=1$, Eq.~(\ref{bezsez11}) giving $5\lambda _1=5\lambda _1$ is trivial, which results in the fact that Eq.~(\ref{bezsez11}) has $M-1$ effective equations).
Moreover, assuming  ${\alpha _{ou{t_{1f}}}}(t) ={\alpha _{ou{t_{1}}}}(t)=E_1{e^{ - {\Gamma}t}}{\sin ^3}{B}t$ (i.e., $\nu_1=\mu_1$), through Eqs.~(\ref{beltabfM}) and (\ref{summary}), we obtain the output single-photon wavepacket  under the Markovian approximation
\begin{equation}
\begin{aligned}
{\alpha _{ou{t_{jf}}}}(t) = {E_1}\sqrt {\frac{{{\gamma _j}}}{{{\gamma _1}}}} {e^{ - \Gamma t}}{\sin ^3}Bt \equiv {E_1}\sqrt {\frac{{{\mu _j}}}{{{\mu _1}}}} {e^{ - \Gamma t}}{\sin ^3}Bt.
\label{bezsez11c}
\end{aligned}
\end{equation}
We point out that the condition of generating any two equal single-photon wavepackets ${\alpha _{ou{t_{jf}}}}(t)$ and ${\alpha _{ou{t_{mf}}}}(t)$ under the Markovian approximation is $\mu _j=\mu _m$ seen from Eq.~(\ref{bezsez11c}), while  the corresponding non-Markovian case for  ${\alpha _{ou{t_{j}}}}(t)={\alpha _{ou{t_{m}}}}(t)$ requires
\begin{equation}
\begin{aligned}
\lambda _j&=\lambda _m,\\
\mu _j &= \mu _m,\\
\nu _j &= \nu _m,
\label{czx1}
\end{aligned}
\end{equation}
which originate from Eqs.~(\ref{summary}), (\ref{alphaoutM}), and (\ref{summary2}).

\begin{figure}
\centering
\includegraphics[height=6.2cm,width=8.5cm]{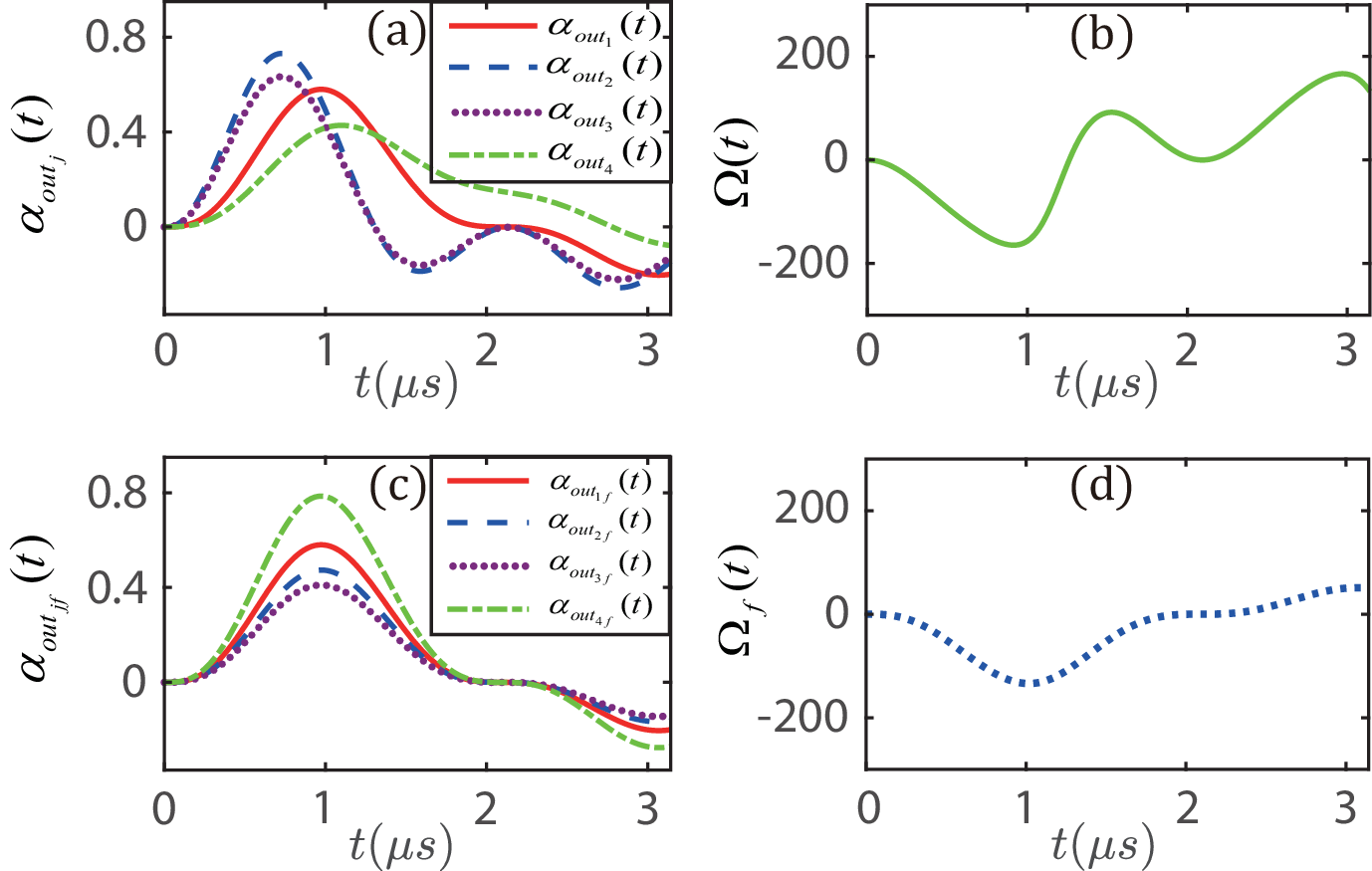}
\caption{(Color online) When the condition in Eq.~(\ref{czx1}) for generating two equal single-photon wavepackets is broken, i.e., $\mu_2=1/6$, $\mu_3=1/8$, $\nu_2=1/3$, and $\nu_3=1/4$ do not satisfy Eq.~(\ref{czx1}), the single-photon wavepackets ${\alpha _{ou{t_2}}}(t)$ and ${\alpha _{ou{t_3}}}(t)$ in the non-Markovian regime are separated in Fig.~\ref{NM_Mar_output_Omegaz2}(a) compared with Fig.~\ref{NM_Mar_output_Omegaz1}(a). In this case, four output single-photon wavepackets ${\alpha _{ou{t_{1f}}}}(t)$, ${\alpha _{ou{t_{2f}}}}(t)$, ${\alpha _{ou{t_{3f}}}}(t)$, and ${\alpha _{ou{t_{4f}}}}(t)$ given by Eq.~(\ref{bezsez11c}) under the Markovian approximation are completely separated in Fig.~\ref{NM_Mar_output_Omegaz2}(c) due to the completely non-equal normalization factors. Eq.~(\ref{bezsez11}) is satisfied by taking the parameters $\mu_1=1/4$, $\mu_4=11/24$, $\nu_1=1/4$, $\nu_4=1/6$, ${\lambda_2} ={\lambda_3} = 24.87$ MHz, ${\lambda_4} = 0.717$ MHz, and the other parameters are the same as Fig.~\ref{NM_output_omega}.}
\label{NM_Mar_output_Omegaz2}
\end{figure}

When all the other spectral widths do not equal the first one shown in Eq.~(\ref{abcz2bb}), by setting $M=4$, we find that four output single-photon wavepackets in  the non-Markovian regime in Fig.~\ref{NM_output_omega}(a) are separated, which originates from the normalization factors $\nu_1=1/4$, $\nu_2=1/5$, $\nu_3=1/2$, $\nu_4=1/20$  and the spectral widths ${\lambda_1} = 2$ MHz, ${\lambda_2} = 1.52$ MHz, ${\lambda_3} = 24.87$ MHz,  ${\lambda_4} = 0.439$ MHz not satisfying Eq.~(\ref{czx1}). In contrast, the output single-photon wavepackets ${\alpha _{ou{t_{1f}}}}(t)$, ${\alpha _{ou{t_{2f}}}}(t)$, ${\alpha _{ou{t_{3f}}}}(t)$, and ${\alpha _{ou{t_{4f}}}}(t)$ under the Markovian approximation are totally overlapped due to the equal normalization factor $\mu_1=\mu_2=\mu_3=\mu_4=1/4$ of Eq.~(\ref{bezsez11c}) in Fig.~\ref{NM_output_omega}(c).
However, in Fig.~\ref{Mar_output_omega}(c), the output single-photon wavepackets $\alpha _{ou{t_{3f}}}(t)$ and $\alpha _{ou{t_{4f}}}(t)$ under the Markovian approximation in Eq.~(\ref{bezsez11c}) are overlapped, which differ from those in the non-Markovian regime in Fig.~\ref{Mar_output_omega}(a), where the partially non-equal normalization factors take $\mu_1=1/4$, $\mu_2=1/10$, $\mu_3=13/40$, and $\mu_4=13/40$.
Moreover, the completely non-equal normalization factors $\mu_1=1/4$, $\mu_2=1/6$, $\mu_3=1/8$, and $\mu_4=11/24$ induce that the multiple output single-photon wavepackets under the Markovian approximation are completely separated in Fig.~\ref{NM_Mar_output_Omega}(c), but they also have different envelops from  those in Fig.~\ref{NM_Mar_output_Omega}(a) in the non-Markovian regime. With the selected parameters meeting condition (\ref{czx1}), two single-photon wavepackets ${\alpha _{ou{t_2}}}(t)$ and ${\alpha _{ou{t_3}}}(t)$ of four single-photon wavepackets in  the non-Markovian regime are equal and  shown in Fig.~\ref{NM_Mar_output_Omegaz1}(a), while Fig.~\ref{NM_Mar_output_Omegaz2} corresponds to the case for all single-photon wavepackets being separated.

To summarize, if the spectral widths of the different environments take the equal values in Eq.~(\ref{abcz2aba}), we show that the output single-photon wavepacket ${{\alpha _{ou{t_j}}}(t)}$ of Eq.~(\ref{beltabM}) in the non-Markovian regime equals  ${{\alpha _{ou{t_{jf}}}}(t)}$ of Eq.~(\ref{beltabfM}) under the Markovian approximation (see Eq.~(\ref{abcz2aa})). However, if all the other spectral widths do not equal the first one in Eq.~(\ref{abcz2bb}), when the first single-photon wavepacket generated by the Markovian system is the same as the non-Markovian case, the same single-photon wavepacket ${{\alpha _{ou{t_j}}}(t)}$ of Eq.~(\ref{beltabM}) in the non-Markovian regime cannot be generated (see Eq.~(\ref{abcz2bbzz})) by ${{\alpha _{ou{t_{jf}}}}(t)}$ of Eq.~(\ref{beltabfM}) under the Markovian approximation because ${{\alpha _{ou{t_{jf}}}}(t)}$ is independent of the spectral width $\lambda_j$. Moreover, we find that the $j$-th output single-photon wavepacket ${\alpha _{ou{t_j}}}(t)$ given by Eq.~(\ref{alphaoutM})
can be expanded as power series of the spectral width $\lambda_j$ as follows
\begin{equation}
\begin{aligned}
{\alpha _{ou{t_j}}}(t) = \sum\limits_{n = 0}^\infty  {{\varepsilon _{j,n}}(t)} \lambda _j^{n + 1},
\label{infiniteseries}
\end{aligned}
\end{equation}
where the time-dependent expansion coefficient ${\varepsilon _{j,n}}(t) = {( - 1)^n}\sqrt {{\gamma _j}} /({\lambda _1}n!\sqrt {{\gamma _1}} ) \cdot \int_0^t {{{(t - {t_1})}^n}} [{\dot \alpha _{ou{t_1}}}({t_1}) + {\lambda _1}{\alpha _{ou{t_1}}}({t_1})]d{t_1}$ is determined when the first output single-photon wavepacket ${\alpha _{ou{t_1}}}(t)$ and decay rate $\gamma_j$ are given. After fixing ${\alpha _{ou{t_1}}}(t)$ and $\gamma_j$, the output single-photon wavepacket ${\alpha _{ou{t_j}}}(t)$ can be controlled by tuning the spectral width $\lambda_j$ (In particular, when $j=1$, Eq.~(\ref{alphaoutM}) or Eq.~(\ref{infiniteseries})   leads to ${\alpha _{ou{t_1}}}(t)={\alpha _{ou{t_1}}}(t)$), which is induced by non-Markovian effects and has no Markovian
counterparts. That is to say, the  system for the multiple single-photon generations in the framework of all the  other spectral widths not equalling the first one in the non-Markovian regime cannot be replaced by the Markovian one, which is the reason we need to consider the non-Markovian system.


Therefore, we point out that these points discussed above may be lost due to making the Markovian approximation when the multiple single-photon wavepackets are generated in the non-Markovian system via setting all the other spectral widths not equalling the first one if the first single-photon wavepackets generated by the Markovian and non-Markovian systems are equal.

\section{EXACT SOLUTIONs FOR THE QUANTUM network DYNAMICS with non-Markovian effects}
\begin{figure*}[htbp!]
\centering
\includegraphics[height=2.6cm,width=15cm]{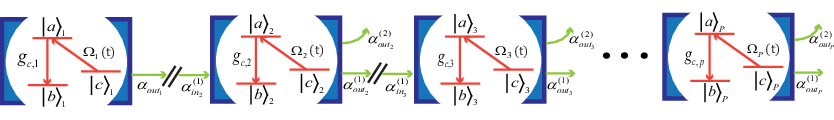}
\caption{(Color online) Illustration of the multiple single-photon generations in driven three-level $\Lambda$ atoms (e.g., cesium atom) coupled to cavities for the non-Markovian quantum network. There are $P$ cavities, where each cavity itself is coupled to the input-output fields forming photonic channels. From the leftmost to rightmost cavities, each contains a driven three-level atom. The coupling parameters and operators defined in the text are indicated. Two cavities are connected by the non-Markovian input-output fields in the following way: the output field of cavity $1$ is directed to cavity $2$ as its input field and so on.}
\label{Ncavity}
\end{figure*}
A non-Markovian quantum network \cite{Zhang201387,Combes20172,Zhang8108,Liu3036,Zhang2017679} is composed of sending and receiving nodes, where the simplest possible configuration of quantum transmission
between two nodes consists of two atoms, which are strongly coupled to their respective cavity modes. Considering non-Markovian input-output fields coupled cavity
chain with $P$ cavities and
each of which contains a driven identical three-level atom (e.g., cesium atom \cite{McKeever2003425,McKeever2004303})
inside in Fig.~\ref{Ncavity}, whose Hamiltonian reads
\begin{small}
\begin{widetext}
\begin{equation}
\begin{aligned}
\hat H =& \sum\limits_{q = 1}^P {\omega _{cav}^{q}} \hat a_q^\dag {{\hat a}_q} + \sum\limits_{q = 1}^P {(\omega _b^q\sigma _{bb}^{(q)} + \omega _c^q\sigma _{cc}^{(q)} + \omega _a^q\sigma _{aa}^{(q)})}  + \sum\limits_{q = 2}^P {\sum\limits_{j = 1}^2 {\int {{\omega _{q,j}}\hat b_{q,j}^\dag ({\omega _{q,j}}){{\hat b}_{q,j}}({\omega _{q,j}})} } } d{\omega _{q,j}} + \int {{\omega _1}\hat b_1^\dag ({\omega _1}){{\hat b}_1}({\omega _1})} d{\omega _1} \\
&+ i\sum\limits_{q = 2}^P {\sum\limits_{j = 1}^2 {\int {d{\omega _{q,j}}[{{\hat a}_q}\hat b_{q,j}^\dag ({\omega _{q,j}}){v _{q,j}}({\omega _{q,j}}) - H.c.]} } }  + i\int {d{\omega _1}[{{\hat a}_1}{{\hat b}_1}({\omega _1}){v _1}({\omega _1}) - H.c.]} {\rm{ }}\\
&+ \sum\limits_{q = 1}^P {({\Omega _q}(t){e^{ - i{\omega _{L}^q}t}}\hat \sigma _{ac}^{(q)} + {g_{c,q}}\hat \sigma _{ab}^{(q)}{{\hat a}_q} + H.c.)} ,
\label{NH}
\end{aligned}
\end{equation}
\end{widetext}
\end{small}
where ${{\hat a}_q}$ is the annihilation operator for $q$th cavity
with frequency $\omega _{cav}^q$, which couples with two non-Markovian input-output fields by the coupling strength ${v _{q,j}}$ (with frequency ${\omega _{q,j}})$ except the first atom-cavity system with ${v _1}({\omega _1})$. $\omega _{b}^q$, $\omega _{c}^q$ and $\omega _{a}^q$ are the frequencies of the ground
state hyperfine levels ${\left| b \right\rangle _q}$, ${\left| c \right\rangle _q}$, and the excited state ${\left| a \right\rangle _q}$ for the atom of the $q$th cavity, respectively.
State ${\left| b \right\rangle _q}$ is coupled to the intermediate ${\left| c \right\rangle _q}$
by the cavity with the coupling strength $g_{c,q}$, and ${\left| c \right\rangle _q}$ is coupled to ${\left| a \right\rangle _q}$ by the driving field ${\Omega _q}(t)$. The atom is initially prepared in state ${\left| c \right\rangle _1}$ in the first cavity, while the atoms of the other cavities are all prepared in state ${\left| b \right\rangle _q}$, and cavities and input fields remain in their vacuum states.
We control the driving field ${\Omega _1}(t)$ to generate an output  single-photon wavepacket
from the first cavity.
By combining the sending and receiving processes, the transfer of
photon from one node to another can be easily accomplished, where the generated photon leaks out of
the first cavity, propagates along the transmission line, and enters
the optical cavity at the second node and so on.
The cavity is coupled with electromagnetic
continuum outside forming a photonic channel \cite{Yao200595,Yao20057}.
The state for the first system can be written as $
\left| {{\Psi }(t)} \right\rangle_1  = {\beta _{b1}}(t)\left| {b,1,0} \right\rangle  + {\beta _{c1}}(t)\left| {c,0,0} \right\rangle  + {\beta _{a1}}(t)\left| {a,0,0} \right\rangle + \int {d{\omega_1} {\alpha _{\omega_1} }(t)\left| {b,0,{1_{\omega_1} }} \right\rangle }.$
But in the $q$th $(q \in [2,P ])$ cavity, the state becomes
\begin{equation}
\begin{aligned}
{\left| {\Psi (t)} \right\rangle _q} &= {\beta _{bq}}(t){\left| {b,1} \right\rangle _q}\left| {{0_1},{0_2}} \right\rangle  + {\beta _{cq}}(t){\left| {c,0} \right\rangle _q}\left| {{0_1},{0_2}} \right\rangle \\
 &+ {\beta _{aq}}(t){\left| {a,0} \right\rangle _q}\left| {{0_1},{0_2}} \right\rangle  + \int {d{\omega _{q,1}}} {\alpha _{{\omega _{q,1}}}}(t)\\
& \cdot {\left| {b,0} \right\rangle _q}\left| {{1_{{\omega _1}}},{0_2}} \right\rangle  + \int {d{\omega _{q,2}}} {\alpha _{{\omega _{q,2}}}}(t){\left| {b,0} \right\rangle _q}\left| {{0_1},{1_{{\omega _2}}}} \right\rangle ,
\label{phi22}
\end{aligned}
\end{equation}
with the initial conditions ${\alpha _{i{n_1}}} = 0$, ${\beta _{b1}}({0}) = 0$, ${\beta _{c1}}({0}) = 1$, and ${\beta _{a1}}({0}) = 0$.
When the photon producing and receiving processes are completed, we have ${\beta _{c1}}({t_1}) = {{\dot \beta }_{c1}}({t_1}) = 0$. If
the initial states of the system in the sending and receiving nodes
are ${\left| c \right\rangle _1}$ and ${\left| b \right\rangle _q}$ $(q \in [2,P ])$, respectively, under
the action of the driving field ${\Omega _q}(t)(q \in [1,P])$, the operation will produce
the entangled states in the sending and receiving nodes by
\begin{widetext}
\begin{equation}
\begin{aligned}
&\left| {{c_1},{b_2},{b_3} \cdots {b_P}} \right\rangle  \otimes \left| 0 \right\rangle \xlongrightarrow{\Omega _1(t)} {\alpha _{{\omega _1}}}\left| {{b_1},{b_2},{b_3} \cdots {b_P}} \right\rangle  \otimes \left| {{\alpha _{ou{t_1}}}} \right\rangle    \xlongrightarrow{\Omega _2(t)} {\alpha _{{\omega _{2,1}}}}\left| {{b_1},{b_2},{b_3} \cdots {b_P}} \right\rangle  \otimes |\alpha _{ou{t_2}}^{(1)}\rangle \\
& + {\alpha _{{\omega _{2,2}}}}\left| {{b_1},{b_2},{b_3} \cdots {b_P}} \right\rangle  \otimes |\alpha _{ou{t_2}}^{(2)}\rangle    \xlongrightarrow{\Omega _3(t)} {\alpha _{{\omega _{3,1}}}}\left| {{b_1},{b_2},{b_3} \cdots {b_P}} \right\rangle  \otimes |\alpha _{ou{t_3}}^{(1)}\rangle  + {\alpha _{{\omega _{3,2}}}}\left| {{b_1},{b_2},{b_3} \cdots {b_P}} \right\rangle \\
& \otimes |\alpha _{ou{t_3}}^{(2)}\rangle  \cdots    \xlongrightarrow{\Omega _P(t)} {\alpha _{{\omega _{P,1}}}}\left| {{b_1},{b_2},{a_3} \cdots {b_P}} \right\rangle  \otimes |\alpha _{ou{t_P}}^{(1)}\rangle  + {\alpha _{{\omega _{P,2}}}}\left| {{b_1},{b_2},{a_3} \cdots {b_P}} \right\rangle  \otimes |\alpha _{ou{t_P}}^{(2)}\rangle,
\label{srchuandi}
\end{aligned}
\end{equation}
\end{widetext}
where $\alpha _{ou{t_q}}^{(1)}(t) = \int {d{\omega _{q,1}}} {\alpha _{{\omega _{q,1}}}}({t_1}){e^{ - i{\Omega _{q,1}}(t - {t_1})}}/\sqrt {2\pi }$ and $\alpha _{ou{t_q}}^{(2)}(t) = \int {d{\omega _{q,2}}} {\alpha _{{\omega _{q,2}}}}({t_1}){e^{ - i{\Omega _{q,2}}(t - {t_1})}}/\sqrt {2\pi }$ with ${\Omega _{q,j}} = {\omega _{q,j}} - {\omega _{cav}^q}$ are the normalized wavepackets of the emitted photon. ${\Omega _q}(t)$ denotes the optimal driving field in the $q$th cavity. In our scheme, there are no interactions between two adjacent cavities (two-sided cavity) \cite{Zhang201387,Zhang8108,Zhang2017679,Liu3036,Yanik200492,Yanik200571}, which can be connected by
the input and output fields.
At the first sending node, there is no incoming photon, i.e., ${\alpha _{in_1}}(t) = 0$, and
we can control the driving field to generate an output  single-photon wavepacket, where the probability amplitudes for the first cavity in the sending node are determined by
\begin{small}
\begin{equation}
\begin{aligned}
&{{\dot \beta }_{b1}}(t) =  - i{g_{{c_1}}}{e^{-i{\delta _{2,1}}t}}{\beta _{a1}}(t) - \int_{0}^t {d\tau {\beta _{b1}}(\tau )} {F_{1,1}}(t - \tau ),\\
&{{\dot \beta }_{c1}}(t) =  - i\Omega _1^*(t){e^{ - i{\delta _{1,1}}t}}{\beta _{a1}}(t),\\
&{{\dot \beta }_{a1}}(t) =  - i{g_{{c_1}}}{e^{i{\delta _{2,1}}t}}{\beta _{b1}}(t) - {\gamma '_1}{\beta _{a1}}(t) - i{\Omega _1}(t){e^{i{\delta _{1,1}}t}}{\beta _{c1}}(t),\\
&{\alpha _{ou{t_{1,1}}}}(t) = \int_{{0}}^t {d\tau {k_{1,1}}(t - \tau ){\beta _{b1}}(\tau )} ,
\label{receivingnode}
\end{aligned}
\end{equation}
\end{small}
where ${\delta _{1,1}} = {\omega _{a}^1} - {\omega _{c}^1} - {\omega _{L}^1}$ and ${\delta _{2,1}} = {\omega _{a}^1} - {\omega _{b}^1} - {\omega _{cav}^1}$ represent
the detunings of the driving field and cavity respectively from atom. ${F_{1,1}}(t) = {\int {| {{v_1}({\omega _1})} |} ^2}{e^{ - i{\Omega _{1,1}}t}}d{\omega _1}$ and ${k_{1,1}}(t) = \int {{v_1}({\omega _1})} {e^{ - i{\Omega _{1,1}}t}}d{\omega _1}$ with ${\Omega _{1,1}} = {\omega _{1}} - {\omega _{cav}^1}$ denote non-Markovian memory and response functions, respectively.

The output field of the first cavity constitutes the input for the second
cavity with an appropriate time delay, i.e., $\alpha _{ou{t_q}}^{(1)}(t - \tau ) = \alpha _{i{n_{q + 1}}}^{(1)}(t)$,
where $\tau $ is a constant related to retardation in the propagation between the mirrors, which is assumed as $\tau=0$ thereafter.
The probability amplitudes with the non-Markovian regime for the receiving node of the $q$th $(q \in [2,P])$ cavity are given by
\begin{small}
\begin{equation}
\begin{aligned}
{{\dot \beta }_{bq}}(t) =&  - i{g_{{c_q}}}{e^{-i{\delta _{2,q}}t}}{\beta _{aq}}(t) - \sum\limits_{j = 1}^2 {\int_{0}^t {d\tau {\beta _{bq}}(\tau )} } {F_{q,j}}(t - \tau )\\
 &+ \sum\limits_{j = 1}^2\int {k_{q,j}^*(t - \tau )} \alpha _{i{n_q}}^{(j)}(\tau )d\tau ,\\
{{\dot \beta }_{cq}}(t) =&  - i\Omega _q^*(t){e^{ - i{\delta _{1,q}}t}}{\beta _{aq}}(t),\\
{{\dot \beta }_{aq}}(t) =&  - i{g_{{c_q}}}{e^{i{\delta _{2,q}}t}}{\beta _{bq}}(t)  - i{\Omega _q}(t){e^{i{\delta _{1,q}}t}}{\beta _{cq}}(t)- {\gamma '_q}{\beta _{aq}}(t),
\label{receivingnode}
\end{aligned}
\end{equation}
\end{small}
where ${\delta _{1,q}} = {\omega _{a}^q} - {\omega _{c}^q} - {\omega _{L}^q}$, ${\delta _{2,q}} = {\omega _{a}^q} - {\omega _{b}^q} - {\omega _{cav}^q}$, ${F_{q,j}}(t) = {\int {| {{v_{q,j}}({\omega _{q,j}})} |} ^2}{e^{ - i{\Omega _{q,j}}t}}d{\omega _{q,j}}$, and ${k_{q,j}}(t) = \int {{v_{q,j}}({\omega _{q,j}})} {e^{ - i{\Omega _{q,j}}t}}d{\omega _{q,j}}$.~The non-Markovian input-output relations can be written as
\begin{equation}
\begin{aligned}
&\alpha _{ou{t_q}}^{(1)}(t)-\alpha _{i{n_q}}^{(1)}(t)  = \int_{{0}}^t {d\tau {k_{q,1}}(t - \tau ){\beta _{bq}}(\tau )} ,\\
&\alpha _{ou{t_q}}^{(2)}(t) = \int_{{0}}^t {d\tau {k_{q,2}}(t - \tau ){\beta _{bq}}(\tau )},\\
&\alpha _{ou{t_q}}^{(1)}(t) = \alpha _{i{n_{q + 1}}}^{(1)}(t),
\label{relations}
\end{aligned}
\end{equation}
where $\alpha _{ou{t_1}}^{(1)}(t) \equiv \alpha _{ou{t_1}}(t)$. In the past, people in general focused on the sending and receiving processes between two cavities \cite{Yao200595,Yao20057}.
Our scheme can happen between multiple cavities, where
each cavity has two input and output fields except the first cavity, where the process of sending and receiving will be repeated all the time.
So as to get the desired wavepackets form or state, we can choose which cavity to end the process. The presented results involving an arbitrary number of driven atom-cavity  might open a way to better understand single-photon generations for non-Markovian quantum networks.

\section{Conclusion}
In summary, we have studied a general control scheme of the quantum system
consisting of $N$ driven three-level atoms coupled to one-side cavity
interacting with the multiple non-Markovian intput-output fields. From the
respect of atoms, there are backflows in the population on the
state $| c \rangle$ in the non-Markovian regime, while there are no
backflows in the Markovian case. Moreover, we calculate the optimal driving
field needed to produce arbitrarily shaped multiple complex single-photon
wavepackets from the cavity in the non-Markovian case, which depends on two
detunings of the cavity and driving field with respect to the three-level
atoms.
Setting all the other spectral widths not equalling the first one results in
that the Markovian system cannot generate the same multiple single-photon wavepackets
as the non-Markovian one when the first single-photon wavepacket generated by the Markovian system is the same as the non-Markovian case, while taking the equal spectral widths values for the different environments can generate this.
The scheme analyzes specifically the exact results of the cavity
interacting simultaneously with the multiple environments in the
non-Markovian regime, where the generated different  single-photon
wavepackets satisfy certain connections with the spectral
parameters. We show that a transition
occurs from non-Markovian to Markovian regimes by controlling the
spectral widths of the environments. Finally, we discuss the dynamics of
quantum network consisting of many cavities containing driven
three-level atoms for non-Markovian intput-output fields.

The studies of non-Markovian multiple single-photon
generations in driven three-level atoms coupled to cavity might
open a way to better understand the multiple single-photon
generations in quantum network and quantum communications.
As an outlook, it will be interesting to explore the multiple
single-photon generations for the total excitation number
non-conserving systems beyond rotating-wave approximations,
e.g., isotropic non-rotating-wave interactions
$\Omega (t){\hat \sigma _{ac}} + {\Omega ^ * }(t){\hat \sigma _{ca}}+
g({\hat \sigma _{ab}} + {\hat \sigma _{ba}})({\hat a} + \hat a^\dag ) $ \cite{is1,is2}
plus $\sum\nolimits_k {{v_k}(\hat a+\hat a^\dag)({{\hat b}_k} +
 \hat b_k^\dag )} $ \cite{shen1,shen2} for case of an atom,
 anisotropic non-rotating wave quantum systems $\sum\nolimits_k [{\alpha _k}
 ({\hat S^\dagger}{{\hat b}_k} + {\hat S}\hat b_k^\dag ) + {\beta _k}
 ({\hat S}{{\hat b}_k} + {\hat S^\dagger}\hat b_k^\dag )]$ with
 $\hat S=\hat \sigma _{ba}$ or $\hat a$ \cite{Tempesta78046608,
 Chen103043708,fan021046,Frhlich79845,Frhlich215291,Nakajima4363,ZLuH},
 and many-body systems \cite{cz1,cz2,cz3,cz4,cz5,cz6},
deserving future investigations.

\section*{ACKNOWLEDGMENTS}

H. Z. Shen would like to thank Dr.~Jia-Xin Yang for valuable  discussions.
This work was supported by National Natural Science Foundation of China under Grants No.~12175033, No.~12274064, No.~12147206, National Key R$\& $D Program of China (No.~2021YFE0193500), and Natural Science Foundation of Jilin Province (subject arrangement project)  under Grant No.~20210101406JC.

\section*{}
\appendix
\section{\label{Hamiltonian1} The derivation of Eq.~(\ref{Htotal})}
The total system is composed of a cavity coupled to $N$ three-level atoms driven by a driving field, where the cavity interacts with $M$ non-Markovian input-output fields in Fig.~\ref{MODEL}, whose Hamiltonian is given by $\hat H' = {\hat H_1} + {\hat H_2}$ with
\begin{small}
\begin{equation}
\begin{aligned}
{\hat H_1} = &{\omega _{cav}}{\hat a^\dag }\hat a + \sum\limits_{m = 1}^N {\left( {{\omega _a}{{\hat \sigma }^{\left( m \right)}}_{aa} + {\omega _b}{{\hat \sigma }^{\left( m \right)}}_{bb} + {\omega _c}{{\hat \sigma }^{\left( m \right)}}_{cc} - i\gamma '{{\hat \sigma }^{\left( m \right)}}_{aa}} \right)},\\
{\hat H_2} = &\sum\limits_{j = 1}^M {\int {{\omega _j}{{\hat b}^\dag }_j\left( {{\omega _j}} \right)} } {\hat b_j}\left( {{\omega _j}} \right)d{\omega _j}\\
&+ \sum\limits_{m = 1}^N {\left[ {\Omega \left( t \right){e^{ - i{\omega _L}t}}{{\hat \sigma }^{\left( m \right)}}_{ac} + {{g}_c}\hat a{{\hat \sigma }^{\left( m \right)}}_{ab} + H.c.} \right]}  \\
&+ i\sum\limits_{j = 1}^M {\int {d{\omega _j}} } \left[ {{\nu _j}\left( {{\omega _j}} \right)\hat a{{\hat b}^\dag }_j\left( {{\omega _j}} \right) - H.c.} \right].
\label{AH0andH}
\end{aligned}
\end{equation}
\end{small}
In a rotating frame defined by ${\hat U} = \exp [ { - i{{\hat H}_1}t} - i{\omega _{cav}}t\sum\nolimits_{j = 1}^M {\int {{{\hat b}^\dag }_j\left( {{\omega _j}} \right)} {{\hat b}_j}\left( {{\omega _j}} \right)}d{\omega _j} ]$ with ${e^{i\omega {{\hat a}^{\rm{\dag }}}\hat at}}\hat a{e^{ - i\omega {{\hat a}^{\rm{\dag }}}\hat at}} = \hat a{e^{ - i\omega t}}$ and ${e^{i{\omega }{{\hat \sigma }_{aa}}t}}{\hat \sigma _{ac}}{e^{ - i{\omega }{{\hat \sigma }_{aa}}t}} = {\hat \sigma _{ac}}{e^{i{\omega }t}}$, we obtain ${\hat H} = {\hat U^{\rm{\dag }}}\hat H'\hat U - i{\hat U^{\rm{\dag }}}\dot {\hat{U}} \equiv \hat H_S + {\hat H}_B + \hat V$ in Eq.~(\ref{Htotal}).

\section{\label{eq15} The derivation of Eq.~(\ref{f})}

Substituting Eq.~(\ref{awjt0}) into Eq.~(\ref{generalEq1}), we have
\begin{equation}
\begin{aligned}
{\dot \beta _b}\left( t \right) =  &- ig_c^ * \sqrt N {e^{ - i{\delta _2}t}}{\beta _a}\left( t \right) \\
&- \sum\limits_{j = 1}^M {\int_{ - \infty }^\infty  {\nu _j^ * } } \left( {{\omega _j}} \right){\alpha _{{\omega _j}}}\left( {{0}} \right){e^{ - i{\Omega _{{\omega _j}}} {t}}}d{\omega _j} \\
& - \sum\limits_{j = 1}^M {\int_0^t {\int_{ - \infty }^\infty  {{{\left| {{\nu _j}\left( {{\omega _j}} \right)} \right|}^2}} } } {e^{ - i{\Omega _{{\omega _j}}}\left( {t - \tau } \right)}}{\beta _b}\left( \tau  \right)d{\omega _j}d\tau .
\label{betabt84}
\end{aligned}
\end{equation}
The first equation of Eq.~(\ref{finallyEq1}) can be derived by substituting Eq.~(\ref{alphainj})-Eq.~(\ref{f}) into Eq.~(\ref{betabt84}).\\

\section{\label{eq11} The derivation of Eq.~(\ref{finallyEq4})}
When ${t_1} \to t $, with Eqs.~(\ref{alphainj}) and~(\ref{alphaoutj}), we obtain
\begin{equation}
\begin{aligned}
{\alpha _{i{n_j}}}(t) + {\alpha _{ou{t_j}}}(t) = & - \frac{1}{{\sqrt {2\pi } }}\int_{ - \infty }^\infty  {{\alpha _{{\omega _j}}}} \left( {{0}} \right){e^{ - i{\Omega _{{\omega _j}}}{t}}}d{\omega _j} \\
&+ \frac{1}{{\sqrt {2\pi } }}\int_{ - \infty }^\infty  {{\alpha _{{\omega _j}}}} \left( t \right)d{\omega _j}.
\label{Dinjsumoutj}
\end{aligned}
\end{equation}
Substituting Eq.~(\ref{awjt0}) into Eq.~(\ref{Dinjsumoutj}) gets
\begin{equation}
\begin{aligned}
{\alpha _{i{n_j}}}(t) + {\alpha _{ou{t_j}}}(t) =& \frac{1}{{\sqrt {2\pi } }}\int_{ - \infty }^\infty  {{\nu _j}\left( {{\omega _j}} \right)\int_{{0}}^t {{\beta _b}\left( \tau  \right)} } \\
&\times {e^{ - i{\Omega _{{\omega _j}}}\left( {t - \tau } \right)}}d\tau d{\omega _j},
\label{Dinjsumoutj2}
\end{aligned}
\end{equation}
which leads to Eq.~(\ref{finallyEq4}) by substituting Eq.~(\ref{kj}) into Eq.~(\ref{Dinjsumoutj2}).\\

\end{document}